\documentclass{mn2e}

\usepackage{graphicx,times,array}

\def \kms {${\rm{km}\,\rm{s}^{-1}}$}

\def \rev  {}

\def \apjs{ApJS}

\def \apj{ApJ}
\def \aj{AJ}
\def \mnras{MNRAS}

 \title[SNELLS low-redshift lenses and the IMF]{The SINFONI Nearby Elliptical Lens Locator Survey:\\Discovery of two new low-redshift strong lenses and 
implications for the initial mass function in giant early-type galaxies\thanks{Based 
on observations collected at the European Southern Observatory, Chile  ESO Programmes 093.B-0193 and 293.B-5026.}}
\author[Russell J. Smith et al.]
{Russell J. Smith$^1$\thanks{Email: russell.smith@durham.ac.uk}, 
John R. Lucey$^1$ and Charlie Conroy$^2$
\\
$^1$ Centre for Extragalactic Astronomy, Department of Physics, Durham University, South Road, Durham \ DH1 3LE \\
$^2$ Harvard Smithsonian Centre for Astrophysics, 60 Garden St., Cambridge, MA 02138, USA
}

\date{Accepted 2015 March 6. Received 2015 February 13; in original form 2014 November 20.}

\pagerange{\pageref{firstpage}$-$\pageref{lastpage}} \pubyear{2015}

\begin{document}

\label{firstpage}

\maketitle

\begin{abstract}
We present results from a blind survey to identify strong gravitational lenses among the population of low-redshift early-type galaxies.
The SINFONI Nearby Elliptical Lens Locator Survey (SNELLS) uses integral-field infrared spectroscopy to search 
for lensed emission line sources behind massive lens candidates at $z$\,$<$\,0.055.
{\rev 
From 27 galaxies observed, we have recovered one previously-known lens (ESO325--G004) at $z$\,=0.034, and 
discovered two new systems, at $z$\,=\,0.031 and $z$\,=\,0.052. 
All three lens galaxies have
high velocity dispersions ($\sigma$\,$>$\,300\,\kms) and $\alpha$-element abundances ([Mg/Fe]\,$>$\,0.3).
From the lensing configurations we derive {\it total} J-band mass-to-light ratios of 1.8$\pm$0.1, 2.1$\pm$0.1 and 1.9$\pm$0.2 within the $\sim$2\,kpc Einstein radius.
Correcting for estimated dark-matter contributions, and comparing to stellar population models 
with a Milky Way (Kroupa) initial mass function (IMF), we  determine the ``mass excess factor'' , $\alpha$.
Assuming the lens galaxies have ``old'' stellar populations (10$\pm$1\,Gyr), the average IMF mass factor
is $\langle\alpha\rangle$\,=\,1.10$\pm$0.08$\pm$0.10, where the first error is random and the second is systematic.
If we instead fit the stellar populations from 6dF optical survey spectra, all three galaxies are consistent with being old, but the age errors
are 3--4\,Gyr, due to limited signal-to-noise. The IMF constraints are therefore looser in this case, with $\langle\alpha\rangle$\,=\,$1.23^{+0.16}_{-0.13}\pm{0.10}$.
Our results are thus consistent with a Kroupa IMF ($\alpha$\,=\,1.00) on average, 
and strongly reject very heavy IMFs with $\alpha$\,$\ga$\,2.
A Salpeter IMF ($\alpha$\,=\,1.55) is inconsistent at the 3.5$\sigma$ level if the galaxies are old, 
but cannot be excluded using age constraints derived from the currently-available optical spectra.}
\end{abstract}   

\renewcommand{\textfraction}{0.1}

\section{Introduction}\label{sec:intro}

Among all techniques for determining galaxy masses, strong gravitational lensing yields the most precise and accurate measurements, with
errors of only a few per cent in favourable cases (Treu 2010; Courteau et al. 2014). With no dependence on the dynamical or thermal state of tracer material, lensing is free
from many of the degeneracies afflicting other mass probes. Being sensitive to all gravitating matter, 
lensing provides some of the strongest direct evidence for dark matter halos around elliptical galaxies. If the contributions of 
dark and baryonic mass can be distinguished, e.g. through their different spatial distribution, then lensing can be used to measure the stellar mass-to-light ratio $\Upsilon$.
Such measurements can provide stringent constraints on stellar population models, and especially on the initial mass function (IMF), since 
faint low-mass stars (and/or dark remnants of high-mass stars) add substantially to the total mass. 

Analysing the lensing properties and stellar velocity dispersions  ($\sigma$) of a sample of 56 strong-lens systems from the SLACS\footnote{Sloan Lenses Advanced
Camera for Surveys.} survey, 
Treu et al. (2010) found that the IMF and the dark-matter halo profile could not {\it both} be universal. Holding the halo profile fixed,
they derived a trend in ``excess'' stellar mass, which they attributed to a non-universal IMF. 
Galaxies with lower velocity dispersion were consistent with Milky-Way-like (e.g. Kroupa 2001 or Chabrier 2003) IMFs, 
while for their most massive galaxies, with $\sigma$\,$\ga$\,300\,\kms, the SLACS modelling
requires stellar mass-to-light ratios twice that expected for a MW-like IMF, i.e. $\alpha$\,=\,$\Upsilon/\Upsilon_{\rm ref}$\,$\approx$\,2 
(where $\Upsilon_{\rm ref}$ is the expected stellar mass to light ratio for a Kroupa IMF, given observational constraints on the star-formation history).
Auger et al. (2010) later showed that heavier-than-MW IMFs were still required on {\it average} even when allowing for contraction of halos in response to the 
stellar component, following Gnedin et al. (2004). With this modification however, the requirement for a {\it trend} in $\alpha$ with galaxy mass is not 
significant, and even the most massive lenses have $\alpha$\,$\approx$\,1.4,
intermediate between Kroupa (2001) ($\alpha$\,=\,1) and a Salpeter IMF\footnote{As usual in extragalactic studies, we (mis)use the term ``Salpeter IMF'' to refer to 
a single-slope power law $N(M)\,\propto\,M^{-x}$ with a slope of $x$\,=\,2.35 over the wide mass range 0.1--100\,M$_\odot$. Salpeter (1955) derived
this slope only over the range 0.5--10\,M$_\odot$.}  ($\alpha$\,=\,1.55). The difference with respect to the Treu et al. result highlights the challenge in separating 
baryonic from dark-matter contributions to the total lensing mass. 

Concurrently with the SLACS analysis, two other largely independent methods also found evidence to favour heavy IMFs in giant ellipticals. 
Fitting dynamical models to integral-field stellar kinematics from the SAURON survey, Cappellari et al. (2006) showed that excess mass
was required in the most massive ellipticals, which could be attributed either to an increasing dark matter fraction or to a systematic trend in the IMF. 
With more detailed modelling of the larger ATLAS3D survey dataset, Cappellari et al. (2012, 2013) 
were able to reject the dark-matter solution, for a range of possible halo prescriptions (see also Thomas et al. 2011). 
They concluded that the IMF mass normalisation indeed increases with velocity dispersion, though the derived trend has a  
shallower slope than that of Treu et al. (2010).  {\rev At 300\,\kms\ the (extrapolated) ATLAS3D trend implies an average $\alpha$\,=\,1.6,
with an estimated intrinsic scatter of 20\,per cent. The two trends are not necessarily inconsistent however, since the ATLAS3D 
sample probes smaller velocity dispersions than the SLACS lenses; Posacki et al. (2015) derived a quadratic $\alpha$-vs-$\sigma$ relation, 
with the slope steepening at high $\sigma$, which adequately describes both samples.
An alternative approach to dynamical modelling using single-fibre spectra for large galaxy samples similarly favours an IMF 
trend reaching  $\alpha$\,$\ga$\,2 at $\sigma$\,$\ga$\,300\,\kms\ (e.g. Dutton et al. 2013; Conroy et al. 2013; Tortora et al. 2014).}

In parallel with these works, van Dokkum \& Conroy (2010) studied gravity-sensitive spectral features in giant ellipticals, and 
found evidence for an excess population of cool dwarf stars, corresponding to a bottom-heavy IMF. The spectroscopic method
is quite distinct from lensing or dynamics, in that the stellar content is probed directly, rather than through any constraint on the total mass, 
and is hence independent of uncertainties due to dark matter. Further studies with this approach (Spiniello et al. 2012; Conroy \& van Dokkum 2012b; 
Smith, Lucey \& Carter 2012; Ferreras et al. 2013; La Barbera et al. 2013; Spiniello et al. 2014) confirmed a trend of increasing dwarf-star content
as a function of galaxy ``mass'' (velocity dispersion or correlated quantities, e.g. element abundances).
Conroy \& van Dokkum (2012b, CvD12b) analysed a sample of $\sim$30 galaxies and derived an implied mass excess (assuming a three-part power-law IMF) 
of $\alpha$\,$\approx$\,2 at $\sigma$\,=\,300\,\kms, similar to the SLACS results. Recent improvements of the models and enlargements to the sample, 
however, yield a wider galaxy-to-galaxy range in $\alpha$ at high mass, and hence a smaller average $\alpha$, but a Milky-Way-like IMF remains disfavoured 
(Conroy \& van Dokkum, in preparation). 

This paper describes an effort to derive robust constraints on $\alpha$ in high velocity dispersion galaxies, by finding and analysing 
strong-lensing galaxies at much lower redshift than those in existing samples.
The accurately determined quantity in a strong-lens system is usually the total projected mass within the Einstein aperture,
defined by a threshold in enclosed surface density (which depends on the angular diameter distances of the lens and source). 
For a given source redshift, the Einstein radius $R_{\rm Ein}$ (in angular units) varies only slowly with lens redshift. Hence, while distant 
lens galaxies (e.g. the ``Cosmic Horseshoe'' studied by Spiniello et al. 2011) have $R_{\rm Ein}$ comparable to, or larger than, the half-light radius 
($R_{\rm eff}$) of the stellar component, in nearby lenses the mass can be determined at a small fraction of $R_{\rm eff}$.
Since stars are more centrally concentrated than dark matter, nearby lenses are optimal for 
measuring the {\it stellar} mass-to-light ratio and hence for constraining the IMF in massive galaxies. A further motivation for studying such systems
is that the lens galaxies can be characterised in great detail, with high signal-to-noise spectroscopy, to 
determine the probable star-formation history, which in turn tightens the limits on $\alpha$.

Only one low-redshift (here defined as $z\la0.05$) massive elliptical strong-lens galaxy has been reported before now\footnote{
The lens of the Einstein Cross system at $z$\,=\,0.04 is a spiral galaxy.
The possible extended arc  around in NGC\,5532 (Hardcastle et al. 2005) has not yet been confirmed with deeper imaging or spectroscopy.
Several {\it cluster-scale}, rather than galaxy-scale, arc candidates have been reported at low redshift (Campusano, Kneib \& Hardy 1998; Blakeslee et al. 2001).
The nearest objects in SLACS are at $z$\,$\sim$\,0.06; they are among the least massive in the sample and do not require heavy IMFs.}.
ESO325--G004 is a $\sigma$\,=\,330\,km\,s$^{-1}$ galaxy at $z$\,=\,0.034
which lenses a $z$\,=\,2.14 background source into a partial Einstein ring with $R_{\rm Ein}$\,=\,2.85\,arcsec (Smith et al. 2005; Smith \& Lucey 2013, hereafter SL13).
Analysing the ESO325--G004 system, SL13 derived a stellar mass-to-light ratio fully consistent with a Milky Way IMF, in apparent contradiction
to the evidence for heavy IMFs in massive galaxies from SLACS, CvD12b and ATLAS3D. 

A key question is whether lensing masses for giant ellipticals {\it in general} favour a MW-like IMF or if ESO325--G004 is simply an
``outlier'' in a broad distribution in $\alpha$. To construct a meaningful {\it sample} of low-redshift strong-lensing early-type galaxies, as required to address this 
issue, we are conducting the SINFONI Nearby Elliptical Lens Locator Survey (SNELLS). Briefly, our approach is to search within infra-red integral-field datacubes,
 to detect background line-emitters behind a carefully-selected sample of candidate lenses. 
In this paper, we give a broad description of the survey and present the first results from SNELLS, including the identification of two new 
multiple-image $z$\,$\la$\,0.05 lens systems, and limits on the IMF in these, plus ESO325--G004, using a uniform dataset and methodology.
Section~\ref{sec:descrip} discusses the motivation for the survey strategy and describes its practical implementation so far. 
Section~\ref{sec:lenses} presents the newly-discovered lenses, and the recovery of the ESO325--G004 system.
In Section~\ref{sec:masses}, we make a first estimate for the lensing masses, mass to light ratios, and the IMF $\alpha$ factor. 
The results are compared to previous work in Section~\ref{sec:disc}, while the conclusions and future prospects
are summarized in  Section~\ref{sec:concs}.

For consistency with other recent work in the field, we adopt cosmological parameters from WMAP7 where necessary, i.e. 
$H_0$\,=\,70.4\,\kms\,Mpc$^{-1}$, $\Omega_{\rm m}$\,=\,0.272 and $\Omega_{\rm \Lambda}$\,=\,0.728 
(Komatsu et al. 2011). We comment in Section~\ref{sec:robalpha} on the 
effect of using instead the Planck Collaboration (2014) cosmological parameters.

\section{A blind survey for lenses among low-redshift ellipticals}\label{sec:descrip}

The two principal obstacles to assembling a sample of low-redshift strong-lensing massive ellipticals are the intrinsic rarity of the lens population
and the low contrast between lens and source in ``typical'' systems. Any search restricted to very bright arcs will have a low yield of lenses per 
candidate, and hence must probe a very large survey volume (e.g. $\sim$7\,Gpc$^3$ within $z$\,$\la$\,0.3 for SLACS).
To find lenses in useful numbers within a ``local'' volume (say 0.07\,Gpc$^3$ within $z$\,$\la$\,0.06), a search must be able to detect systems in
which the lensed source is very faint, a task made more difficult by superposition on a bright foreground galaxy. Traditional methods
struggle to identify lenses in this regime. 

As we show in this section, targetted integral field unit (IFU) observations, especially in the infra-red (IR), can provide the improved contrast necessary 
to discover less visually-spectacular, but more numerous, lens systems at low redshift.

\subsection{IFU lens-search strategy}

\begin{table*}
\caption{Summary of the 27 SNELLS target galaxies observed in ESO Period 93. Redshifts, $z$, and velocity dispersions, $\sigma$ are from
the 6dF Peculiar Velocity Survey (Campbell et al. 2014) or from SDSS DR7 (Abazajian et al. 2009) as indicated.
}\label{tab:p93obs}
\begin{tabular}{llcccll}
\hline
Target & NED ID & source & $z$ & $\sigma$ [\kms] & Lens ID & Notes \\
\hline
J0058513--162809 & NGC 0333B  				& 6dF &  0.0541 & $313\pm20$ & \\
J0141423--073528 & 2MASX J01414232--0735281 	& 6dF &  0.0519 & $320\pm18$ & SNL-2 \\
J0145535--065608 & IC 0158  					& 6dF &  0.0523 & $328\pm20$ & \\
J0202308--505553 & ESO197--G018 			& 6dF &  0.0215 & $333\pm24$ & \\
J0213108--062026 & MCG--01--06--071 			& 6dF &  0.0433 & $308\pm21$ & \\
J0258177--044906 & MCG--01--08--023  			& 6dF &  0.0308 & $324\pm27$ & SNL-3  \\
J0302383--603219 & 2MASX J03023835-6032198   & 6dF &  0.0543 & $332\pm26$ & \\
J0420410--144651 & IC 0367 					& 6dF &  0.0308 & $303\pm20$ & \\
J0837138--032141 & CGCG 004--087 			& 6dF &  0.0402 & $313\pm16$ & & J only  \\
J0857060--055418 & MCG--01--23--010 NED01	& 6dF &  0.0454 & $313\pm16$ & & J only  \\
J1153082--323357 & ESO440--G025 			& 6dF &  0.0271 & $313\pm22$ & \\
J1343331--381033 & ESO325--G004 			& 6dF &  0.0339 & $356\pm16$ & SNL-0 & H only \\
J1352252--345600 & ESO384--G007 			& 6dF &  0.0382 & $356\pm21$ & \\
J1352533--282921 & NGC 5328 				& 6dF &  0.0158 & $335\pm\phantom{0}8$   & \\
J1450280+110649 & 2MASX J14502809+1106497 	& SDSS &  0.0533 & $323\pm\phantom{0}6$   & \\ 
J1510556--112847 & NGC 5872 				& 6dF &  0.0249 & $329\pm14$ & \\
J1511314+071507 &  CGCG 049--033 			& SDSS &  0.0446 & $334\pm\phantom{0}6$   & \\ 
J1557496+161836 & NGC 6023 				& SDSS &  0.0370 &$328\pm\phantom{0}5$   & \\ 
J1601361+122136 & CGCG 079--022 			& SDSS &  0.0356 & $325\pm\phantom{0}5$   & \\
J1613129+174828 & 2MASX J16131301+1748283  & SDSS &  0.0372 & $316\pm\phantom{0}7$   & \\ 
J1832293--601726 & 2MASX J18322937--6017262  & 6dF &  0.0517 & $358\pm46$ & \\
J1916325--401233 & NGC 6768 				& 6dF &  0.0186 & $310\pm13$ & \\
J1958358--523509 & FAIRALL 0877  			& 6dF &  0.0509 & $307\pm24$ & \\
J2100286--422523 & ESO286--G022 			& 6dF &  0.0312 & $356\pm18$ & SNL-1 \\
J2129374--211144 & IC 1386 					& 6dF &  0.0356 & $319\pm14$ & \\
J2318463--102357 & IC 1479 					& 6dF &  0.0317 & $317\pm19$ & \\
J2324259+135820 & NGC 7651 				& SDSS & 0.0435 & $322\pm\phantom{0}5$   & \\ 
\hline
\end{tabular}
\end{table*}

Previous optical searches for galaxy-scale strong lenses fall mainly into two categories,
each with its strengths and limitations: 

{\it Spectroscopic} approaches, like SLACS (Bolton et al. 2006) 
look for anomalous emission lines from background galaxies in integrated spectra from large surveys (see also Warren et al. 1996; Hall et al. 2000).
Because the spectrum is obtained from a large aperture (a 3$\arcsec$ fibre for the Sloan Digital Sky Survey, SDSS), 
only very bright-lined sources can be detected over the dominant continuum flux from the lens galaxy. 
In SLACS, moreover, the SDSS spectral range imposes a limit on the source redshift ($z_{\rm src}$\,$\la$\,1),
and the search volume behind each target is correspondingly small. After spectroscopic selection,
follow-up data is needed to establish the image positions, either using {\it HST} to heighten contrast against the 
lens galaxy, or with an IFU to map the line emission.

{\it Morphological} methods (e.g. Cabanac et al. 2007) instead search systematically for arc-like features in broad-band images, either using automated
algorithms (e.g. Belokurov et al. 2009; Gavazzi et al. 2014) or visual inspection (including crowd-sourced classification, e.g. Marshall, Lintott \& Fletcher 2014). 
When lensed images are found, follow-up spectroscopy is necessary to establish the source and lens redshifts, before
physical properties can be determined. With ground-based data, this approach becomes less effective for detecting low-$z$ lenses, 
where the arcs lie inside the effective radius of the lens galaxy and are hard to distinguish from stellar shells, spiral arms, tidal features, etc.

The rarity of high-contrast lenses can be overcome by applying either the spectroscopic or the morphological search techniques to large
(and pre-existing) survey datasets, but most of the resulting lenses are then inevitably found at moderate redshifts. Neither method is well suited 
to detecting faint arcs behind nearby massive galaxies.

ESO325--G004 was serendipitously identified as a lens through inspection of
{\it Hubble Space Telescope} images (Smith et al. 2005). In this case, the necessary contrast between lens and source is provided by the 
high angular resolution of the images, and by the much bluer colour of the arcs compared to the elliptical. 
Perhaps surprisingly, archival {\it HST} data is not presently available for many very massive nearby ellipticals, except for  
those at the centres of rich clusters, which are poor candidates for {\it galaxy-scale} lenses (see Section~\ref{sec:samp}). 
The discovery of ESO325--G004 suggests that a systematic blind search with {\it HST} for arcs behind candidate lenses might
be fruitful, but it would be an observationally expensive programme, and still require spectroscopic follow up for redshift measurements.

For the lens search described here, we instead use ground-based IFU spectroscopy to combine some of the advantages
of both the spectroscopic and imaging-based methods. The IFU data concentrate the emission spatially as well as in wavelength, 
increasing contrast  over the lens galaxy ($\ga$40$\times$ lower lens galaxy flux in a 1$\arcsec$  
aperture at 0.25$\times$$R_{\rm eff}$, compared to a full 3$\arcsec$ SDSS fibre). To maximize the background volume
available per lens candidate, we work in the near IR; this also allows us to target the bright H$\alpha$ emission line. 
A combination of J and H bands probes lensed H$\alpha$ over a volume-per-candidate $\sim$5 times larger than that
available to SLACS (mainly lensed [O{\,\sc iii}] emitters in optical spectra).
Finally, since we perform dedicated observations, rather than exploiting an existing survey, we maximize efficiency by pre-selecting
a sample of the most promising lens candidates based on velocity dispersion. The lensing cross-section scales as $\sigma^4$, for 
isothermal mass profiles, giving a factor of 20 gain at $\langle\sigma\rangle=315$\,\kms\ compared to $\langle\sigma_{\rm SDSS}\rangle=150$\,\kms.
Together, these factors yield a search efficiency which is $\ga$1000 times larger, on a per target basis, than the SLACS approach, 
so that a useful sample of lenses can be assembled from observations of only a few tens of candidates.

\begin{figure*}
\includegraphics[angle=270,width=175mm]{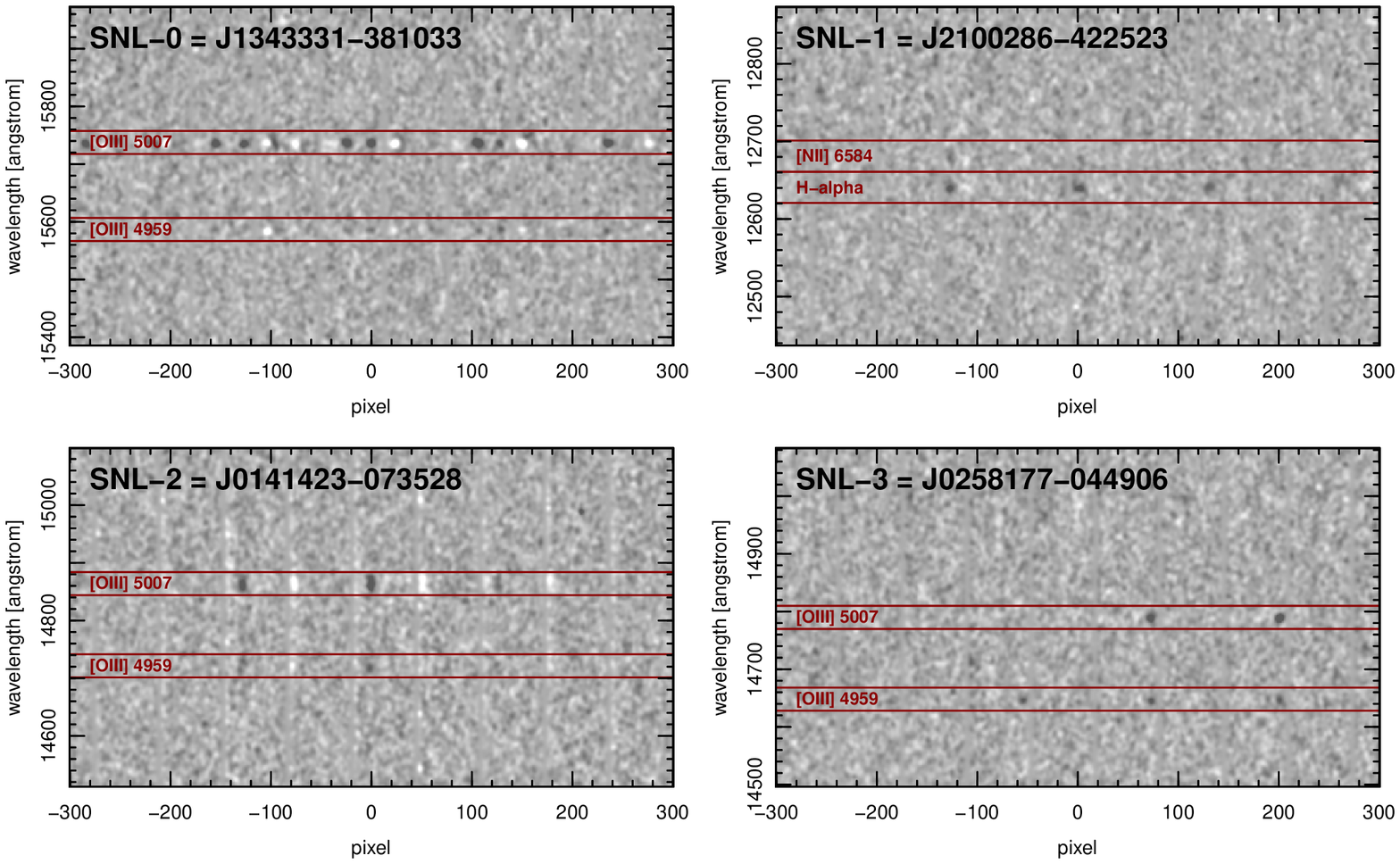}
\caption{Sections from the ``detection images'' showing detected background line emission in the four galaxies described in Section~\ref{sec:lenses}.
The detection image is a wavelength-calibrated, noise-normalised unwrapped cube, i.e. with the IFU slices placed alongside each other (the slice edges are faintly
visible in SNL-2). The two pointings have been subtracted, resulting in positive and negative signals, and processed to suppress signal from the 
lens galaxy and residual sky, highlighting the background emission line features.}
\label{fig:inspects}
\end{figure*}

\subsection{Implementation}

The specific implementation of our approach uses the SINFONI instrument (Eisenhauer et al. 2003; Bonnet et al. 2004), 
a near-IR image-slicing IFU spectrograph at the ESO 8.2m Very Large Telescope (VLT), Unit Telescope 4. 

We established the optimal survey design based on the following assumptions:
\begin{enumerate}
\item{The lenses have singular isothermal ellipsoid profiles, with $\sigma_{\rm SIE}$\,=\,315\,\kms, $z_{\rm lens}$\,=\,0.04 and ellipticity of 
0.3. For this model, the cross-section for multiple-image lensing is 25\,arcsec$^2$ for a distant source ($z_{\rm src}$\,$\ga$\,0.5).}
\item{The sources are H$\alpha$ emitters drawn from the evolving luminosity function as determined from wide-field narrow-band imaging
by Sobral et al. (2013).  We adopt a fixed faint-end slope $\alpha_{{\rm H}\alpha}=-1.6$, a linear relationship for the break luminosity
$\log L^*_{{\rm H}\alpha}$ with $z$, and interpolate the evolution of the luminosity density, $\log \phi^*_{{\rm H}\alpha}(z)$, with a cubic spline.}
\item{The line detection limit is computed as a function of wavelength using the ESO exposure time calculator,  including the 
effects of instrument efficiency, sky emission lines, atmospheric absorption, etc. The total survey yield is computed by integrating the counts
down to the detection limit and summing over the whole wavelength coverage of the observation.
}
\end{enumerate}
In principle, lensing magnification increases the total source flux, but initial {\it detection} of peaks will not be aided fully by this 
factor, because the signal is also spread into more pixels. Since detectability is mainly governed by surface brightness, 
we conservatively ignored the magnification effect in our calculations.

Experimenting with the parameters of this calculation shows that for a given {\it total} exposure time, the yield is maximized by splitting
the observation between J and H bands, to increase the background volume surveyed. Likewise, the slope of the source counts function is such that 
shallow observations of more targets are preferable to deep observations of fewer candidates.
The final predicted survey yield, i.e. identified lenses per observed candidate, is $\sim$10\,per cent, for integration times of $\sim$1\,hr split between the J and H bands. 
Hence, we calculated that a survey of only $\sim$30-40 galaxies is needed to yield a meaningful sample of 3--4 clean IMF-constraining 
low-$z$ lens systems, and compiled a target sample accordingly.

\subsubsection{Sample}\label{sec:samp}

To maximize the number of detectable lensed sources, the candidate sample is explicitly biased towards the most massive galaxies.
Targets were selected from the Six-degree Field (6dF) Peculiar Velocity Survey (Campbell et al. 2014), which covers the whole southern sky 
with $|b|>10^\circ$, and from the Sloan Digital Sky Survey (SDSS, Data Release 7, Abazajian et al. 2009)  in the north, with R.A. and declination cuts 
suitable for observation in the April--September period from the VLT.

The primary selection was on stellar velocity dispersion, 
with $\sigma_{\rm 6dF}$\,$>$\,300\,km\,s$^{-1}$ or  $\sigma_{\rm SDSS}$\,$>$\,310\,km\,s$^{-1}$ (the difference
accounts for the different fibre sizes in the two surveys, 6.7\,arcsec vs 3.0\,arcsec, given typical velocity dispersion gradients). 
Campbell et al. measured velocity dispersions only for 6dF galaxies with $z$\,$<$\,0.055; we apply the same redshift limit on the SDSS candidates. 
A signal-to-noise limit (15\,\AA$^{-1}$) was imposed to exclude galaxies with large errors on velocity dispersion. 
These initial selection criteria produced a list of $\sim$130 candidates.

In clusters or high-mass galaxy groups, the large dark-matter component can push the lensing critical curve to large radius. 
Such cases, in which lensing measures the cluster-scale halo, rather than the galaxy mass, are much less suitable for constraining the 
stellar IMF. We used the NASA Extragalactic Database, together with compilations of group velocity dispersion ($\sigma_{\rm grp}$) data 
(e.g. Struble \& Rood 1999; Ragone et al. 2006; Crook et al. 2007; Coziol et al. 2009) to exclude central galaxies of 
groups with $\sigma_{\rm grp}$\,$>$\,400\,\kms). 
Additionally, we removed non-central galaxies that lie within (or projected close to) very massive clusters, to improve robustness of the lensing masses.
Targets from 6dF at low galactic latitude ($|b|<20^\circ$) were also excluded, since cluster/group catalogues are less complete in these regions.

All available survey images of the targets were inspected to remove galaxies with double nuclei or very close companions,
which might have influenced the measured velocity dispersion (e.g. Bernardi et al. 2006). The SDSS and 6dF survey spectra were inspected to remove galaxies
with strong emission lines, since the presence of young stars, or continuum from an active galactic nucleus, would hinder the eventual inference on the IMF.

Of the 38 galaxies reaching this stage in the selection process, only four have been imaged with {\it HST}. One of these is the known lens 
ESO325--G004, which was retained in the sample (for H-band observation only) as a method verification target. 
Another object with HST data is NGC\,5532, where a faint arc-like feature has been reported from shallow ultra-violet imaging (Hardcastle et al. 2005), but has not been 
confirmed with spectroscopy; we retained this candidate in the target list (although it was not ultimately observed). 
For the final two galaxies (NGC 6768 and ESO508--IG045), the {\it HST} images do not suggest the presence of any lensed background galaxies;
these targets were removed from the SINFONI sample.
The final target list hence comprises 36 lens candidates.

\begin{figure*}
\vskip 4mm
\includegraphics[angle=270,width=175mm]{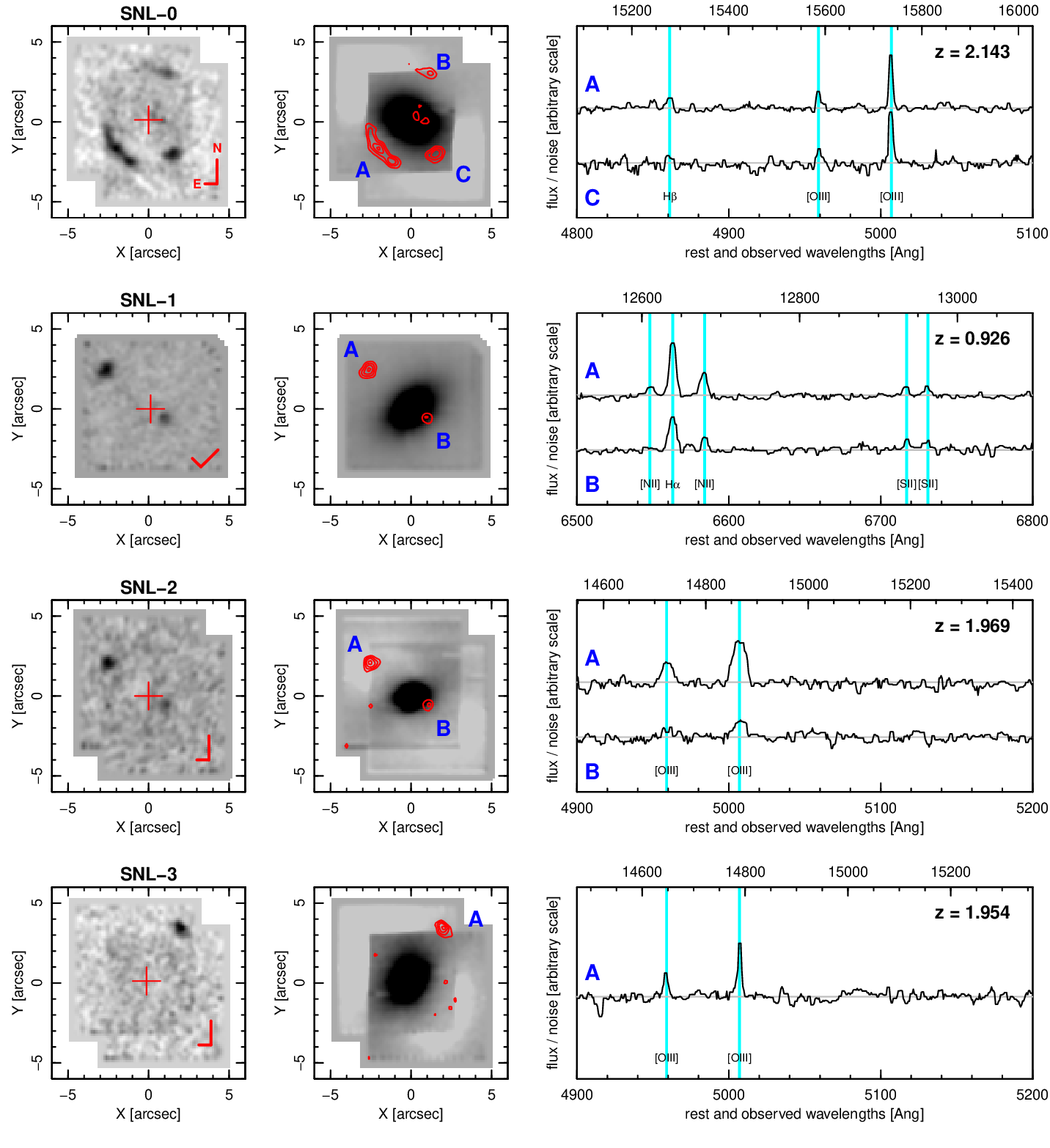}
\caption{Source and lens images, and extracted spectra, for the SNELLS lens candidates. For each source, the first panel shows a narrow-band, continuum-subtracted
image extracted around the brightest line (H$\alpha$ for SNL-1, [O\,{\sc iii}] 5007\,\AA\ in all others). 
The lens galaxy centre is marked by the red cross; the compass shows the field orientation. The second panel shows the broad-band image
as greyscale, with contours indicating the narrow-band flux, and source images identified. The third panel shows spectra extracted from the source images. These
spectra have been normalised by the errors to suppress sky subtraction residuals, hence they are strictly closer to signal-to-noise spectra. 
The strongest emission lines are indicated.
For SNL-0 and SNL-3, 
the data used are the original SNELLS observations (two offset exposures). For SNL-2, two repeated observations (four exposures) were combined. 
For SNL-1, we show the deeper follow-up data obtained in director's discretionary time.}
\label{fig:sinfofig}
\end{figure*}

\begin{figure*}
\includegraphics[angle=270,width=180mm]{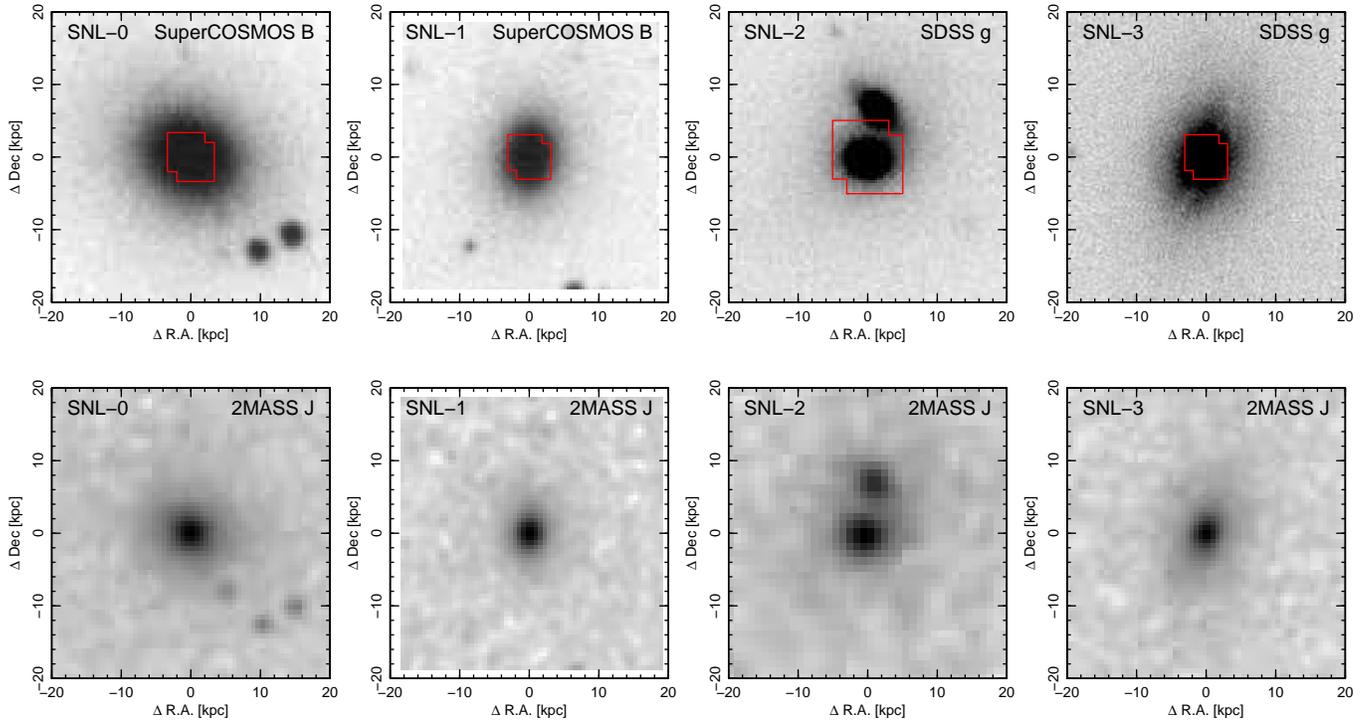}
\vskip -2mm
\caption{Images of the three confirmed SNELLS lenses SNL-0,1,2, and the candidate SNL-3. The upper row shows optical
imaging for each galaxy, from SDSS where available, and otherwise from SuperCOSMOS digitized Schmidt plates (Hambly et al. 2001).
The red outline indicates the $\sim$10$\times$10\,arcsec$^2$  footprint of the SNELLS observations.
The lower row shows the 2MASS J-band images (Jarrett et al. 2000).}
\label{fig:candimg}
\end{figure*}

\begin{figure*}
\includegraphics[angle=270,width=180mm]{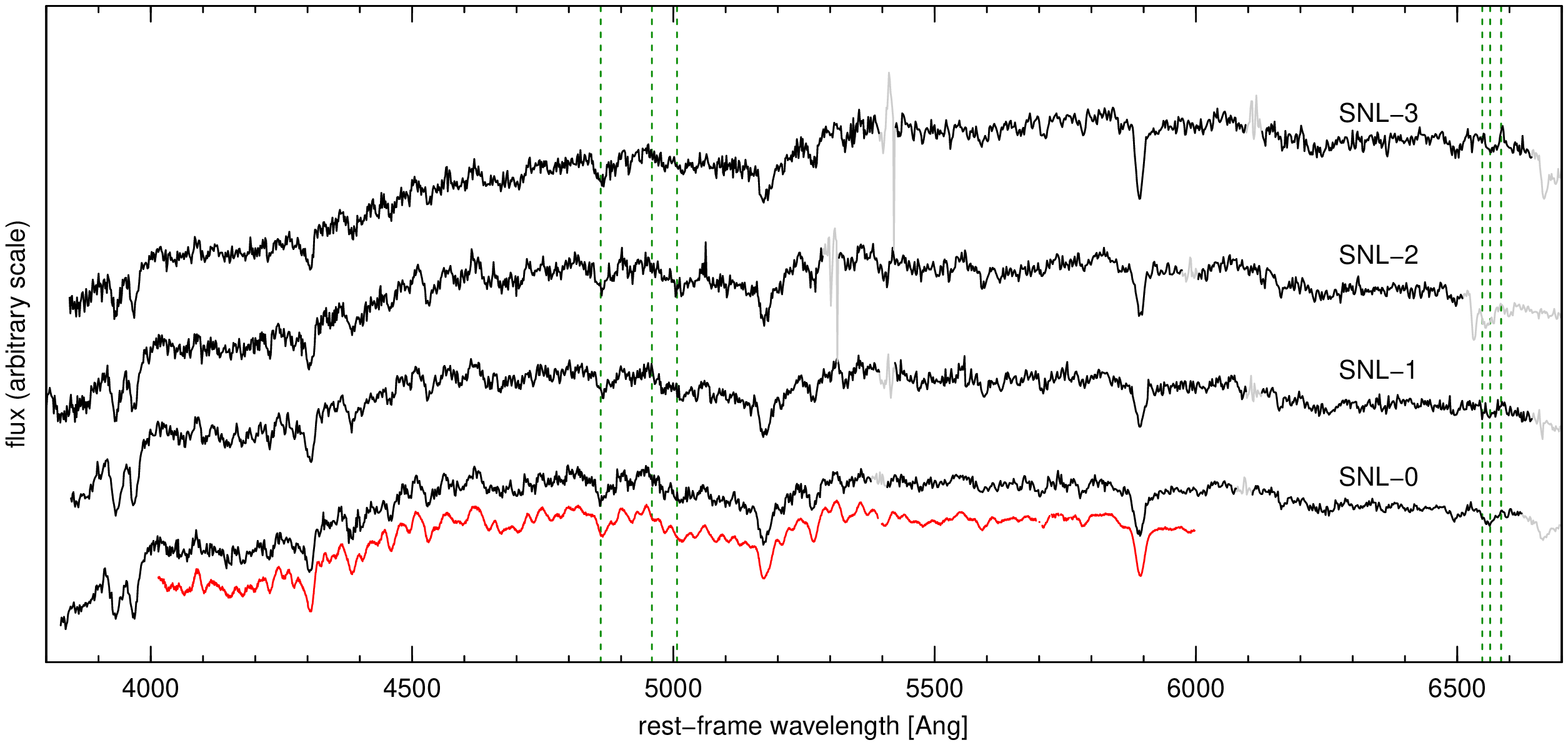}
\vskip -5mm
\caption{Optical spectra of the three SNELLS lenses SNL-0,1,2, and the candidate SNL-3, from the 6dF Galaxy Survey (Jones et al. 2004, 2009). 
The signal-to-noise ratio in each spectrum is 20--30\,\AA$^{-1}$.
For SNL-0 we also show the high-quality VIMOS spectrum analysed in SL13, with signal-to-noise $\sim$400\,\AA$^{-1}$ (red). 
Grey sections indicate spectral regions affected by sky lines or atmospheric absorption.
Green dotted lines indicate the potential location of 
H$\alpha$, H$\beta$, [O\,{\sc iii}] and [N\,{\sc ii}] emission lines. Weak [N\,{\sc ii}] emission is seen in SNL-3, but 
there is no evidence for any nebular emission in SNL-0,1,2.}
\label{fig:candspec}
\end{figure*}

\subsubsection{Observations and data reduction}

The SINFONI observations were made in service mode during ESO Period 93 (April--September 2014). 
A complete observation for each target consists of two pointings in J band and two in H band, with a single 600-sec exposure in each pointing, per band. 
Hence the total exposure time for each galaxy is 40 minutes (rather than the 60 minutes assumed in the the survey yield estimate).
Using the largest pixel scale available with SINFONI, each pointing covers a field-of-view 8$\times$8\,arcsec$^2$, sampled with 0.250$\times$0.125\,arcsec$^2$ pixels.
The two pointings were offset by 2.3\,arcsec, always aligned NE--SW. 
The spectral resolution is $\sim$2000 in J and $\sim$3000 in H. 
At the close of Period 93, a total of 27 candidate galaxies had been observed, from the original sample of 36. 
Two candidates were observed only in J band due to technical problems, while the known lens ESO325--G004 was targetted for observation only in H.
The complete list of observed targets is summarized in Table~\ref{tab:p93obs}.
 
Each pair of offset exposures was reduced using standard methods from the ESO SINFONI pipeline, 
to produce an ``unwrapped cube'', in which all spatial slices are rebinned onto a calibrated wavelength scale, but formatted as a two-dimensional spectrum,
containing positive and negative traces of the target, together with residual airglow signal. No correction for telluric absorption was made.
Further processing steps specific to SNELLS were applied to remove the residual sky and the
strong continuum from the elliptical galaxy, leaving any faint background emission lines untouched. 
The galaxy signal was  removed by 
modelling the source as having a spatially-invariable spectrum (derived from the bright central region), 
modulated by a low-order continuum correction fitted to each spatial pixel. 
The median residual sky was constructed and subtracted for each spatial slice. 
After masking bad pixels (based on a Laplace transform filter) and the extreme ends of each slice, 
all pixels were normalised by the marginal rms function computed in 
both the spatial and the spectral direction, to yield an approximation to a signal-to-noise image, which was then smoothed to highlight features on 
the scale expected for emission-line features. 
We refer to this product, still in unwrapped cube format, as the ``detection image''. For most galaxies
this is essentially a blank frame with fairly uniform noise. 

Candidate lenses were identified through visual inspection of the detection images to locate background emission lines\footnote{
Although a well-designed automated detection algorithm might be able to locate fainter sources, such systems would probably
require dedicated deep follow-up observations to confirm them as lenses, and hence lack one of the advantages of the SNELLS approach.}.
The characteristic features are peaks on the appropriate 
spatial and spectral scale, usually present both in positive and negative (i.e. in both two spatial pointings), extending across multiple image slices, and not 
coincident with the galaxy nucleus. Clear signals of this form were found in four cases, as shown in Figure~\ref{fig:inspects}. 
We present details of these systems in the next section.

 \section{Identified lenses}\label{sec:lenses}

After identifying emission lines in the detection image, we extract narrow-band images centred on the brightest line, to establish the spatial form of the system,
and compute spectra for the brightest images to determine the source redshifts. 
Figure~\ref{fig:sinfofig} shows the SNELLS data for each galaxy, in the form of (a) the 
continuum-subtracted narrow-band image centred on the brightest detected line from the source,
(b) a broadband image showing the lens, with the narrow-band contours overlain, and (c) the extracted spectra from the brightest images of the source. 

In three of the four candidates (one of which is ESO325--G004), we recover multiple emission-line sources with identical wavelengths; the interpretation of these systems
as strong gravitational lenses is compelling. In the fourth case, only a single source image is observed. 

For additional context, we show the available optical/IR imaging for the lens candidates in Figure~\ref{fig:candimg} and the optical spectra from 6dF in 
Figure~\ref{fig:candspec}. The following sections describe each candidate lens in turn.

\subsection{J1343331--381033 (ESO325--G004, SNL-0)}

This system is the previously-known low-redshift lensing elliptical (Smith et al. 2005; SL13), which we assign the name SNL-0 for the purposes of this paper. 
Satisfying our SNELLS selection criteria, this galaxy was observed only in the H band, in order to test methods for 
data processing and source detection, in a case where the presence of lensed emission lines is known in advance.
SNL-0 is the central galaxy of a poor group (Abell S0740), and has boxy outer isophotes but no disturbances in its inner regions. 
Compared to the distribution of sizes and luminosities of high-$\sigma$ galaxies, SNL-0 has a large effective radius (6.6\,kpc, compared to 
median 5\,kpc for $\sigma$\,$>$\,300\,\kms\ galaxies from Campbell et al. 2014) and large total luminosity (4$\times$$10^{11}L_\odot$, compared
to median 2$\times$$10^{11}L_\odot$.). The 6dF optical spectrum (supported by a high-S/N VIMOS spectrum from SL13)
shows an apparently passive stellar population, with weak Balmer absorption lines and no traces of emission. 

The SNELLS data for SNL-0 are shown in the top row of Figure~\ref{fig:sinfofig} .
The narrow-band emission-line image clearly recovers most of the structure of the arc system as seen in HST images, though some fidelity is lost from
combining only two offset frames. Our observations recover the  [O{\sc iii}] doublet and weak H$\beta$ line, as previously seen in long-slit data from SL13.
No other lines are detected.

\subsection{J2100286--422523 (ESO286--G022, SNL-1)}

This is the first newly-discovered lens system from SNELLS (Figure~\ref{fig:sinfofig}, row 2). 
With an effective radius of only 2\,kpc, SNL-1 is three times smaller than SNL-0, and among the most compact objects among the  
$\sigma$\,$>$\,300\,\kms\ galaxies from Campbell et al. (2014). 
The total luminosity of SNL-1 is accordingly smaller ($\sim$$10^{11}\,L_\odot$).
Due to its southerly declination, SNL-1 has not been covered by any modern public optical imaging survey; the SuperCOSMOS and 2MASS  
images show an elongated but otherwise featureless morphology (Figure~\ref{fig:candimg}). 
The 6dF spectrum suggests an old passive stellar population, with weak H$\gamma$ and H$\beta$ absorption,
and no emission detectable at H$\alpha$ or elsewhere (Figure~\ref{fig:candspec}). There are no other galaxies or groups close to SNL-1;
in fact, this object is included in the catalogue of unusually isolated galaxies compiled by Karachentseva et al. (2010). 

The original SNELLS data revealed a background emission line in the J-band, forming a source at 3.7\,arcsec from the target centre and 
a suspected faint counter image on the opposite side of the target, at 1.1\,arcsec.
Deeper J-band follow-up observations of SNL-1 were obtained in a director's discretionary time (DDT) observation with SINFONI on 2014 August 12.
We observed for a total of 60 minutes on-source, with equal time on fully-offset background areas to provide a cleaner sky subtraction.
Three different detector orientations were used, to help in rejecting cosmetic defects from the images.
The deep and clean follow-up data confirm the strongest detected line as H$\alpha$  at a redshift of $z$\,=\,0.926.
The [N\,{\sc ii}]\,6548,\,6584\,\AA\  and [S\,{\sc ii}]\,6717,\,6731\,\AA\ lines are also detected in the brighter image. 
The fainter counter-image is unambiguously detected in the DDT observation, with multiple lines
confirming the lensing interpretation beyond reasonable doubt. 

With $z_{\rm lens}$\,=\,0.031, SNL-1 is now the closest confirmed galaxy-scale strong lens.

\subsection{J0141423--073528 (2MASX J01414232--0735281, SNL-2)}

This is our second newly-discovered lens system (Figure~\ref{fig:sinfofig}, row 3). 
SNL-2, with $z_{\rm lens}$\,=\,0.052, is part of a galaxy pair  (Figure~\ref{fig:candimg}); the companion galaxy is $\sim$1\,mag fainter, located 7\,arcsec to the North. 
SDSS imaging shows that the pair is embedded in a common, asymmetric stellar halo, indicating the 
galaxies are at a common distance and suggesting a recent or ongoing tidal interaction. SNL-2 itself appears very regular; its effective radius (6.0\,kpc)
is comparable to SNL-0, and its total luminosity is only slightly smaller (3$\times$$10^{11}L_\odot$). 
The 6dF spectrum is consistent with a pure passive stellar continuum (Figure~\ref{fig:candspec}), but 
the H$\alpha$/[N{\sc ii}] region is contaminated by telluric absorption so we cannot rule out weak nebular emission as securely as for
SNL-0 and SNL-1.

By chance, two separate SNELLS observations were obtained for SNL-2, so we combine the data to improve total depth\footnote{No observatory flux standard was observed for the first observation, so the target was returned to the queue.}. The H-band observation reveals emission in two lines consistent with 
the [O\,{\sc iii}]\,4959,\,5007\,\AA\  doublet, at a redshift of  $z$\,=\,1.97. (The observer's-frame wavelength is $\sim$1.48\,$\mu$m, placing the lines at the short-wavelength
edge of the H band, in a region of poor atmospheric transmission.)
The lines are noticeably broad, with FWHM $\sim$550\,\kms.
The emission is concentrated in a bright source at 3.2\,arcsec from the galaxy centre, with a 
fainter second image at radius 1.2\,arcsec diametrically opposite. 
The 4959-\AA\ line is not clearly seen in the fainter image, but we consider the concordant 5007-\AA\ line, and the symmetric configuration, 
sufficient to confirm this as a lensed counter-image. No counterpart to the emission-line source is visible in the SDSS imaging.

\subsection{J0258177--044906 (MCG--01--08--023, SNL-3)}

Finally, we report a further candidate lens system, which cannot be modelled sufficiently well to include in our mass estimates  (Figure~\ref{fig:sinfofig}, row 4). 
In SNL-3, the lens is an edge-on S0 galaxy at $z$\,=\,0.031,  with a prominent dust lane seen in SDSS imaging (Figure~\ref{fig:candimg}). The 6dF spectrum shows weak 
emission at [N\,{\sc ii}]\,6584\,\AA\ from the lens candidate itself (Figure~\ref{fig:candspec}).

The SNELLS data 
reveal emission lines consistent with the [O\,{\sc iii}]\,4959,\,5007\,\AA\ 
doublet at a redshift of $z$\,=\,1.95 (as for SNL-2, this places the lines in a region of low atmospheric transmission at the edge of the H-band). 
The source is 3.9\,arcsec from the candidate lens centre and appears to be 
tangentially extended; however, no counter-image can be identified with the present data. No counterpart to the emission-line source is visible in the SDSS imaging. 

There are three possible interpretations for SNL-3: (a) the background galaxy is not located in the multiply-imaged region in the source plane, in which case only a 
weak constraint on the lensing mass can be derived; (b) the source is doubly imaged, with a faint inner counter-image that is too faint to detect without deep follow-up; (c) 
an outer counter-image lies beyond the SINFONI observational footprint, implying a very large lensing mass (we discuss field-of-view selection bias more 
generally in Section~\ref{sec:fovbias}).

Although improved data might eventually distinguish between these possibilities, the complexity of the lens galaxy (dust lane, emission lines)
makes this a poor system for inferring the IMF mass factor in any case, and no further observations are planned for SNL-3.

\begin{figure}
\includegraphics[angle=270,width=85mm]{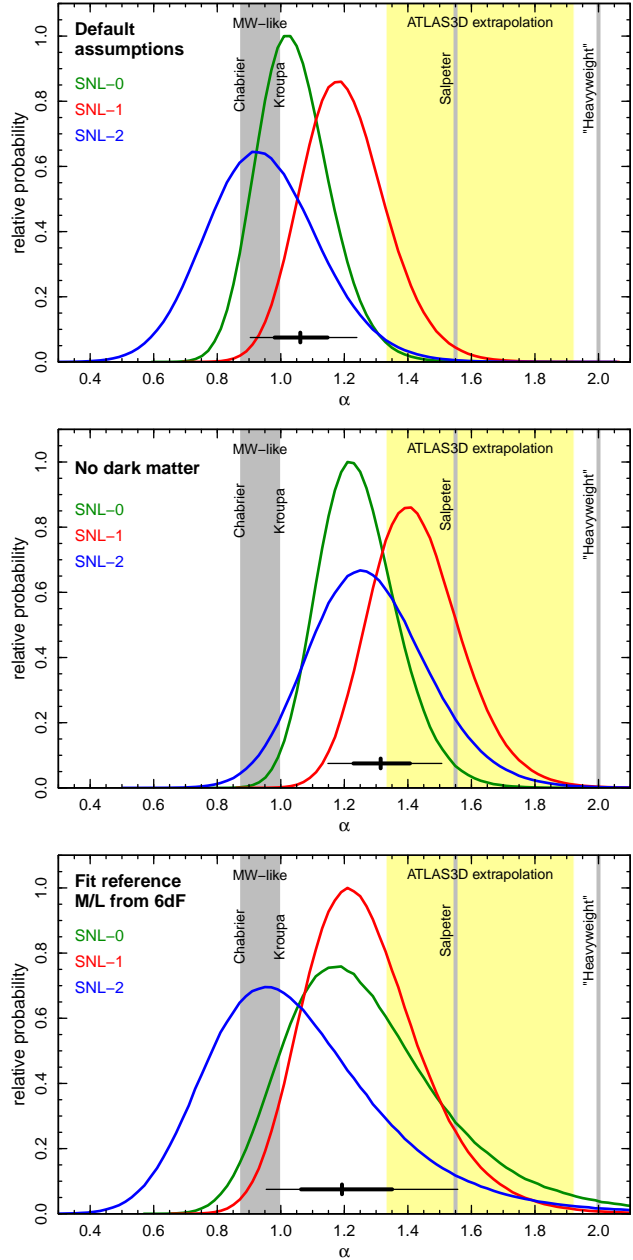}
\caption{IMF mass factor constraints for the three SNELLS lens galaxies. In the upper panel, the heavy solid curves show the probability distribution for $\alpha$ in
each galaxy, adopting our default assumptions (old stellar population, EAGLE dark matter contributions). The second panel shows the results under the extreme 
assumption of negligible dark matter within the Einstein radius. The third panel shows the results obtained when fitting $\Upsilon_{\rm ref}$ from the 6dF spectra, 
rather than assuming the populations are old. 
The heavy/light bar below shows the 1$\sigma$/2$\sigma$ intervals for the mean of the three galaxies. For comparison, the yellow shading indicates the $\pm$1$\sigma$ in $\alpha$ 
expected at the mean velocity dispersion of our sample, 345\,\kms, based on the trend reported by Cappellari et al. (2013) (slightly extrapolated since their 
sample contains no very high-$\sigma$ galaxies). The $\alpha$ values for several candidate IMFs are shown with grey bars. Here, ``heavyweight'' refers to $\alpha$\,=\,2, 
corresponding to the heaviest results from Treu et al. (2010) and Conroy \& van Dokkum (2012b).}
\label{fig:alphapdf}
\end{figure}

\renewcommand{\arraystretch}{1.5}

\begin{table*}
\caption{Observed quantities, corrections and derived parameters for the three confirmed SNELLS lenses.}\label{tab:obspars}
\begin{tabular}{lcccl}
\hline
Name & SNL-0 & SNL-1 & SNL-2 & notes \\
\hline
Target 						& J1343331--381033 & J2100286--422523 & J0141423--073528  \\
NED ID						& ESO325--G004  & ESO286--G022 & 2MASX J01414232--0735281 \\
\hline
$z_{\rm lens}$  					& 0.034 & 0.031 & 0.052  & lens redshift \\
$z_{\rm src}$ 					& 2.141 &  0.926 & 1.969 & source redshift \\
$D_L$ [Mpc] 					& 149 & 135 & 230  & luminosity distance to lens\\
$D_A$ [Mpc] 					& 139 & 127 & 208  & angular diameter distance to lens\\
$f = D_{\rm s}/D_{\rm ls}$			& 1.027 & 1.043 & 1.044 & lensing geometry factor\\
$D$ [Mpc] = $f D_A$				& 143 & 132 & 217 & effective distance to lens \\ 
$\Sigma_{\rm crit}$ [$10^3M_\odot$\,pc$^{-2}$]  & 12.31 & 13.05 &  7.35 & lensing critical surface density\\ 
\hline
$\sigma_{\rm 6dF}$ [\kms]			& 356$\pm$16 & 356$\pm$18 & 320$\pm$18 & aperture corrected to $R_{\rm eff}/8$\\
$[$Mg/Fe$]$						& 0.33$\pm$0.05 & 0.31$\pm$0.05 & 0.38$\pm$0.06 & full-spectrum fit to 6dF \\
$R_{\rm Eff}$ [arcsec] (J)		 	& 9.8 & 3.3 & 5.9 &  PSF-corrected\\
$r_{\rm Eff}$ [kpc] (J)		 	& 6.6 & 2.0 & 6.0 &  PSF-corrected\\
$L_{\rm Tot}$  [$10^{10}L_{\odot,J}$]	 & 40.9 &  9.8 & 25.5 \\ 
\hline
$R_{\rm Ein}$ [arcsec]			& 2.85 & 2.38 & 2.21 & 0.5$\times$image-separation for SNL-1,2\\
$r_{\rm Ein}$ [kpc]  				& 1.92 & 1.47 & 2.23 \\
$f_{\rm corr}$				 	& 1.00 & 0.96 & 0.88 &  correction factor for shear / asymmetry \\  
$M_{\rm Ein}$ [$10^{10}M_\odot$] 	& 14.3$\pm$0.7 & 8.9$\pm$0.4 & 11.5$\pm$1.2   \\ 
\hline
$J_{\rm Ein}$ 				 	& 12.36 & 12.80 & 13.53 & PSF-corrected\\ 
$A_J$ 						& 0.04 & 0.03 & 0.02 & Schlafly \& Finkbeiner (2011) \\
$k_J$ 						& --0.02 & --0.02 & --0.03 & bandshifting  correction (via {\tt EzGal})\\
$L_{\rm Ein}$ [$10^{10}L_{\odot,J}$]	& 7.8$\pm$0.2 & 4.3$\pm$0.1 & 6.2$\pm0.2$  & assumes $M_{\odot,J}$\,=\,3.71 as in Maraston (2005)\\
\hline
$M/L$ [solar units, J-band]         	& 1.82$\pm$0.10 & 2.08$\pm$0.12 & 1.87$\pm$0.22 & total, i.e. includes DM \\ 
\hline
\end{tabular}
\end{table*}

\section{From lensing to IMF constraints}\label{sec:masses}

The observed lensed can be used to constrain the IMF mass excess factor, $\alpha$, defined by
\[
\alpha = \frac{\Upsilon}{\Upsilon_{\rm ref}} = \left(\frac{M_{\rm Ein} - M_{\rm DM}}{L_{\rm Ein}}\right)  \cdot \frac{1}{ \Upsilon_{\rm ref}}  
\]
where 
	$M_{\rm Ein}$ is the total lensing mass within the Einstein radius,
	$M_{\rm DM}$ is the contribution from dark matter, 
	$L_{\rm Ein}$ is the Einstein aperture luminosity,
	and $\Upsilon_{\rm ref}$ is a reference mass-to-light ratio for the stellar population assuming a fiducial IMF.

In this section we discuss the measurements and assumptions made for each quantity above, 
and combine them to yield constraints on $\alpha$.
The lensing mass and the associated luminosity are strongly constrained by the data (Section~\ref{sec:lensml}), 
while $M_{\rm DM}$ and $\Upsilon_{\rm ref}$ involve assumptions 
which are either not determined at all from the observations (dark matter) or only partly informed by data (stellar population model).
Our default treatment of these quantities is described in Section~\ref{sec:defalpha}. In Section~\ref{sec:robalpha} we test alternative
assumptions and discuss other possible systematic effects on our results. Finally, in Section~\ref{sec:fovbias}, we discuss a selection
bias due to the limited field of view of the IFU observations.

\subsection{Lensing M/L}\label{sec:lensml}

In this section we report the derivation of the total projected mass and projected luminosity within the Einstein radius of each lens.

For SNL-0, we simply adopt the angular Einstein radius, $R_{\rm Ein}$\,=\,2.85\,arcsec, as determined by Smith et al. (2005) from fitting HST imaging
using a singular isothermal sphere mass model, with a linear external shear term. 

For SNL-1 and SNL-2, we approximate $R_{\rm Ein}$ by half the image separation, and compute the
associated lensing mass using the symmetric lens-mass formula, 
\[
M_{\rm Ein} = \frac{c^2 R_{\rm Ein}^2}{4G}  D_{\rm l}  \frac{D_{\rm s}}{ D_{\rm ls}} = 1.23\times10^{10} M_\odot \left ( \frac{R_{\rm Ein}}{1^{\prime\prime}} \right )^2 \left ( \frac{D}{100\,{\rm Mpc}} \right )
\]
(e.g. Refsdal \& Surdej 1994). The effective distance $D$ is calculated from the angular diameter 
distances $D_{\rm l}$ (observer to lens), $D_{\rm s}$ (observer to source) and  $D_{\rm ls}$ (lens to source).
{\rev It is implicit in this approach that the lensing potential is assumed to be fixed on the centre of light; this is clearly justified
if the lensing mass is indeed dominated by stars.}

Using the {\tt gravlens} code (Keeton 2001), we have verified that the mass projected within a circular aperture, with radius
given by the half-separation of the image pair, is well reproduced by applying the symmetric approximation, even when the image configuration is 
quite far from symmetric, as in SNL-1 and SNL-2.
For isothermal and de Vaucouleurs profile mass models with ellipticities up to $e$\,=\,0.25, we find a scatter of $\la$3\,per cent, and a bias $\la$1.5\,per cent.
External shear introduces a comparable level of noise. Including all sources of error, we estimate uncertainties of $\sim$5\,per cent on the lensing masses.

For both SNL-1 and SNL-2, we apply small corrections to the symmetric lens approximation. 
In SNL-1, the light distribution has clear ellipticity and the images are aligned along the minor axis. 
In this configuration, if the lensing mass is distributed with similar shape and
orientation as the light (as must be the case if stellar mass dominates within $R_{\rm Ein}$), then the lensing mass is overestimated by applying
the symmetric lens approximation. For isothermal and de Vaucouleurs  profiles with ellipticities $e$\,=\,0.2--0.3, we find that $M_{\rm Ein}$ should be revised
downwards by 4\,per cent for SNL-1.

For SNL-2, there is an additional complication due to the close neighbouring galaxy, which must induce some distortion to the observed lens configuration.
We explore the probable influence of the companion using {\tt gravlens}. We build a two-component model, with singular isothermal sphere mass profiles fixed 
at the coordinates of SNL-2 and the neighbouring galaxy 7\,arcsec north. With increasing mass contributed by the neighbour, the source-plane (unlensed) location
of the emission line object must move northwards, and the normalization of the primary lens must be reduced, in order to maintain the same observed 
image configuration. The true mass projected within the ``naive'' $R_{\rm Ein}$ (the half-separation of the images) is also smaller if we allow for substantial mass 
in the companion. Hence the effect of neglecting the external shear would be to overestimate the lensing mass of SNL-2. 
If the companion has 10--50\,per cent of the mass of the primary lens
(as suggested by the flux ratio), then the lensing mass for SNL-2 should be revised downwards by 8--16\,per cent. We adopt a correction of 12\,per cent and
an increased lensing mass error of 10\,per cent to account for greater uncertainty in the external shear for this system.

With these corrections applied, the Einstein masses for SNL-0, SNL-1 and SNL-2 are 14.3, 8.9 and 11.5 $\times$$10^{10}\,M_\odot$, respectively,  
with adopted errors of 5, 5, and 10\,per cent.

To derive the Einstein aperture luminosity $L_{\rm Ein}$ on a common basis for all three lenses (two of which -- SNL-0 and SNL-1 -- 
are beyond the coverage of SDSS), we use J-band photometry from 2MASS (Jarrett et al. 2000). The J band is preferred over H or K for
this purpose, 
because the fluxes will ultimately be compared to stellar population models, which show better agreement among themselves at J
than at longer wavelengths (see e.g. fig 4 of Mancone \& Gonzalez 2012). 

Aperture fluxes were measured directly from the image data within the Einstein radius of each lens. A correction for the point-spread function (PSF) 
was derived by fitting a convolved S\'ersic model using {\tt galfit} (Peng et al. 2010). The convolution kernel was derived from stars in the same image frame as the 
target galaxy. The PSF corrections are $\sim$0.25\,mag, and we estimate them to be accurate to 0.02\,mag. 
{\rev This uncertainty is added in quadrature with photometric errors tabulated in the 2MASS database which, for such bright galaxies, are dominated by systematic
calibration errors of $\sim$0.02--0.03\,mag.}
Small corrections are applied for band-shifting (``k-correction''), based on estimates from various models implemented in {\tt EzGal}
(Mancone \& Gonzalez 2012), and for galactic extinction, from Schlafly \& Finkbeiner (2011). 
To convert from flux to luminosity, we use the Hubble-flow luminosity distance in the adopted cosmology, and the absolute magnitude
of the Sun $M_J$\,=\,3.71, as assumed in the Maraston (2005) models that we use to estimate $\Upsilon_{\rm ref}$ below.
The resulting $L_{\rm Ein}$ are 7.9, 4.3 and 6.1 $\times$$10^{10}L_\odot$  for SNL-0, SNL-1, and SNL-2 respectively, {\rev with total estimated errors of 3--4}\,per cent.
The total J-band mass-to-light ratios (including dark matter) for the three galaxies are therefore 1.8$\pm$0.1, 2.1$\pm0.1$ and 1.9$\pm$0.2.

\subsection{IMF normalisation factor}\label{sec:defalpha}

\begin{table*}
\caption{Conversion from total lensing mass-to-light ratio to IMF mass factor $\alpha$. We
summarize the limits and/or assumptions made for the projected dark-matter mass (within the Einstein radius), $M_{\rm DM}$, and for the 
reference mass-to-light value, $\Upsilon_{\rm ref}$, and report the resulting constraints on $\alpha$.  
Part (a) shows the assumptions and results for our ``default'' treatment (see Section~\ref{sec:defalpha}), while subsequent parts show 
how the results are affected by changes to our assumptions and methods (Section~\ref{sec:robalpha}). {\rev The final column reports the mean $\alpha$
for the three lenses; including a systematic error component of 10\,per cent.}
}\label{tab:alpha} 
\begin{tabular}{lllccccc}
\hline
 & Assumptions & Quantity  & SNL-0 & SNL-1  & SNL-2 & $\langle\alpha\rangle$ \\
\hline
(a) & Eagle DM & $M_{\rm DM}$ [$10^{10}M_\odot$]   & $2.28_{-0.32}^{+0.38}$ & $1.39_{-0.20}^{+0.24}$ & $2.98_{-0.42}^{+0.49}$ \\
& Assumed old (M05 models) & $\Upsilon_{\rm ref}$	   & 1.47$\pm$0.12 & 1.47$\pm$0.12 & 1.47$\pm$0.12 &  \\
& (default result) & $\alpha = \Upsilon/\Upsilon_{\rm ref}$ 	 &  1.04$\pm$0.11 & 1.20$\pm0.13$ & 0.94$\pm$0.17 & 1.06$\pm$0.08$\pm$0.10 \\
\hline
(b) & SL13 dark matter method &  $M_{\rm DM}$ [$10^{10}M_\odot$]   &  1.86$^{+1.37}_{-0.79}$  &  1.15$^{+0.85}_{-0.49}$ & 2.57$^{+1.90}_{-1.09}$ \\
& & $\alpha = \Upsilon/\Upsilon_{\rm ref}$   & 1.07$^{+0.13}_{-0.16}$ & 1.23$^{+0.15}_{-0.19}$ & 0.99$^{+0.20}_{-0.27}$ & $1.10_{-0.12}^{+0.11}$$\pm$0.10 \\
\hline 
(c) & No dark matter & $\alpha = \Upsilon/\Upsilon_{\rm ref}$ 	& 1.24$^{+0.13}_{-0.11}$ & 1.42$^{+0.15}_{-0.13}$ & 1.27$^{+0.19}_{-0.17}$ &   1.31$\pm$0.09$\pm$0.10 \\
\hline
(d) & Fit ages from 6dF (CvD12 models) &  $\Upsilon_{\rm ref}$	   & $1.27\pm0.23$ & $1.46\pm0.18$ & $1.40\pm0.27$ & \\ 
& & $\alpha = \Upsilon/\Upsilon_{\rm ref}$ &  1.25$^{+0.29}_{-0.21}$ & 1.25$^{+0.19}_{-0.16}$ & 1.03$^{+0.30}_{-0.23}$ & 1.19$^{+0.16}_{-0.13}$$\pm$0.10 \\ 
\hline
(e) & Allow for FoV selection bias & &  & & & 1.10$\pm$0.08$\pm$0.10 \\
\hline
\end{tabular}
\end{table*}

In this section we describe the derivation of the IMF factor, $\alpha$, using our ``default'' assumptions both for  the 
dark matter content and for the stellar population model.

The dark-matter corrections within $R_{\rm Ein}$ are expected to be small, since the lensing measurement
probes the baryon-dominated inner part of the galaxy. In SL13, we estimated the correction for SNL-0 
based on halo properties in pure dark-matter simulations. We assumed Navarro, Frenk \& White (1996) halo profiles, 
with the mass--concentration relation of Neto et al. (2007), and imposing limits
on the halo mass according to the observed velocity dispersion of the surrounding group.
{\rev The limitations of the approach adopted in SL13 are (i) that the simulations do not trace any response
of the halo the to the growing presence of a dense stellar system at its centre 
and (ii) that the calculation assumes the scatter in the  mass--concentration relation is uncorrelated
with the likelihood of hosting a giant elliptical galaxy\footnote{\rev Neto et al. (2007) show that 
much of the scatter around the concentration--mass relation is attributable to variation in halo formation redshift. 
If ellipticals are preferentially found in the earliest-forming 
halos, the SL13 method would underestimate the dark-matter content on average, and overestimate the galaxy-to-galaxy scatter.}.}

For the present work, we instead use dark matter masses estimated directly from the EAGLE\footnote{Evolution and Assembly of GaLaxies and their Environments.}
cosmological hydrodynamical simulation (Schaye et al. 2015). This simulation includes state-of-the-art baryonic physics prescriptions 
and is able to reproduce a wide range of observables for the local galaxy population. Enclosed (i.e. 3D) density profiles from EAGLE have been
presented as a function of halo mass by Schaller et al. (2015), who showed that the inclusion of baryons significantly affects the central regions.
To ensure that the corrections are as well matched as possible, Schaller kindly recomputed {\it projected}
profiles for galaxies for galaxies that match the selection of lens candidates in SNELLS. Specifically, we selected simulated halo-central 
galaxies with stellar mass $10^{11}$--$10^{12}M_\odot$ and projected stellar velocity dispersions $\sigma$\,$>$\,275\,\kms, 
but excluding halos with total mass greater than $10^{14}M_\odot$ (to reject massive groups and clusters).
With this selection, we find that the projected dark matter masses from EAGLE, interpolated to match the Einstein radii for our lenses, are remarkably similar 
among galaxies. The mean EAGLE $M_{\rm DM}$ measured at the Einstein radius for SNL-0, SNL-1 and SNL-2 is 2.3, 1.4 and 3.0$\times$$10^{10}M_\odot$ 
respectively, with a halo-to-halo scatter of only 17\,per cent in each case. Expressed as a fraction of the total lensing mass, the 
dark-matter correction is $16^{+3}_{-2}$\,per cent for SNL-0 and SNL-1, and $26^{+5}_{-4}$\,per cent for SNL-2.

Other possible non-stellar mass contributions within the Einstein radius, e.g. gas and central black holes, are neglected. 
We implicitly assume that all gas lost during stellar evolution has either been heated and dispersed into a low-density halo or else recycled into 
subsequent generations of stars, and hence does not remain as an additional mass component to be subtracted. 
The average central black-hole mass for $\sigma$\,=\,300\,\kms\ early-type 
galaxies is $\sim$1.4$\times$$10^9\,M_\odot$ (McConnell et al. 2011), an order of magnitude smaller than the dark-matter contribution.

The reference stellar mass-to-light ratio, $\Upsilon_{\rm ref}$, is the ratio expected for each galaxy assuming a universal fiducial IMF, here taken
as that of Kroupa (2001). The $\Upsilon_{\rm ref}$ depends mainly on galaxy ``age'', in a luminosity-weighted sense, with a weak effect from metallicity.
In SL13, we fitted to Lick indices measured on a high-S/N spectrum of SNL-0, to determine age and metallicity constraints, from which we derived
limits on $\Upsilon_{\rm ref}$. For the enlarged sample of three lenses, we do not have equivalently high-quality spectra for 
all of the galaxies. For the ``default'' results of this paper, we instead proceed by assuming all three galaxies are ``old'', 
{\rev which is supported by analysis of larger galaxy samples (e.g. McDermid et al. 2015).}
Specifically, we adopt  $\Upsilon_{\rm ref}$ values appropriate for a mean age of 10\,Gyr, and allow a standard deviation of 1\,Gyr. 
The corresponding 2$\sigma$ range for formation redshift is 1\,$<$\,$z$\,$<$\,4.
 At solar metallicity, the predicted distribution for (rest-frame J-band) $\Upsilon_{\rm ref}$ from Maraston (2005) models has mean 
1.47 and standard deviation 0.12. A factor of two in metallicity yields only a 3\,per cent change in $\Upsilon_{\rm ref}$ in J.

Finally, 
we combine the above quantities and their probability distributions to arrive at estimates for $\alpha$, relative to the Kroupa IMF: 
$\alpha$ = 1.04$\pm$0.11, 1.20$\pm$0.13 and 0.94$\pm$0.17, for SNL-0, SNL-1 and SNL-2, respectively.  
The larger error for SNL-2 reflects the larger dark-matter correction necessary due to its larger distance, and also the more uncertain
shear due to the companion galaxy. The result for SNL-0 is consistent with that obtained in SL13 (1.04$\pm$0.15), despite several 
changes to the assumptions, input data and methodology. 

The principal result of this paper, then, is that all three known low-redshift early-type lens galaxies have mass-to-light ratios consistent with 
a Kroupa IMF.  Each individually is incompatible with a Salpeter ($\alpha$\,=\,1.55) or heavier IMF, at the $>$99.5\,per cent level, with respect 
to the random uncertainties (we discuss systematic errors below). The average $\alpha$ for these three very massive ($\sigma$\,$>$\,300\,\kms) 
early-type galaxies is $\langle\alpha\rangle$\,=\,1.06$\pm$0.08. 
This result is summarized in Part (a) of Table~\ref{tab:alpha}, and in the upper panel of Figure~\ref{fig:alphapdf}.

\subsection{Robustness and systematics}\label{sec:robalpha}

According to the default assumptions made in the previous section, our results favour lightweight IMFs in all three confirmed SNELLS lenses.
In this section we explore the sensitivity of the result to modifications in our methodology and inputs. 

The lensing mass and corresponding luminosity, both projected quantities within the nominal Einstein radius, are the most robustly
determined properties of the systems. Neither quantity could be systematically in error at the $\sim$50\,per cent level necessary to derive a Salpeter IMF. 
Instead, the main uncertainties arise in correcting the total lensing mass to a stellar mass, and in referencing the stellar mass-to-light ratio to the 
value expected for a fiducial IMF. 

The dark-matter mass contribution is an uncertain input to the calculation, as it is not estimated directly from the data. 
However, since the correction factor is small, plausible variations in the dark matter treatment make only small impact on $\alpha$. 
To illustrate this, we have repeated the calculation using estimates based on the concentration--mass relation in pure dark-matter simulations, 
as in SL13. In this case, the mean projected dark-matter masses are $\sim$20\,per cent smaller than in the default model 
The SL13 scheme also yields much larger scatter in dark-matter content, and hence a slightly 
increased error in $\alpha$. Adopting this approach in place of the EAGLE scheme leads to $\langle\alpha\rangle$\,=\,$1.10_{-0.12}^{+0.11}$ 
 (Table~\ref{tab:alpha}, part b). Even the making the limiting assumption of negligible dark matter within the Einstein radius, 
 the average derived $\alpha$ would increase only to $\langle\alpha\rangle$\,=\,1.31$\pm$0.09 (see Table~\ref{tab:alpha}, part c, 
 and the second panel of Figure~\ref{fig:alphapdf}). 
 
A second input that is not yet derived from the data is the assumption that all of the lens galaxies have uniformly old stellar populations. 
Younger galaxies have lower mass-to-light ratios for a given IMF, and hence assuming more recent star formation would yield a larger estimate 
of $\alpha$. In order to favour a Salpeter IMF, we would need $\Upsilon_{\rm ref}$\,$\approx$\,1.0, implying ages 
$\sim$6\,Gyr, {\it on average}, for the three galaxies. While late-epoch star-formation is known to occur in some early-type galaxies
(evidenced by emission lines, strong Balmer absorption, central dust lanes or ring/spiral structures) this is rarer among those with very high velocity dispersion. There
is no detectable emission in the 6dF spectra for SNL-0, SNL-1 or SNL-2, nor do any of them visually show strong Balmer absorption lines
suggestive of recent star formation, i.e. bursts within the past $\sim$1\,Gyr. Intermediate stellar ages, however, cannot be so easily excluded.

As an alternative to assuming old ages, we have also derived probability distributions for $\Upsilon_{\rm ref}$ based on fitting the 
6dF spectra over the 4000--5400\,\AA\ wavelength range, using the latest version of the Conroy \& van Dokkum (2012a, CvD12a) stellar population models. 
Given the limited signal-to-noise (20--30\,\AA$^{-1}$), we fit a single burst stellar population with a restricted set of element abundances. 
Confidence intervals on $\Upsilon_{\rm ref}$ were derived through a Markov Chain Monte Carlo sampling method. 
Note that the 6dF spectra were obtained through a fibre of diameter 6.7\,arcsec, and hence include some light from beyond the Einstein radius, but the 
mass-to-light ratio from the integrated aperture is unlikely to be significantly biased by this mismatch. 
From the spectral fitting approach fits we derive J-band $\Upsilon_{\rm ref}$, assuming a 
Kroupa IMF, of 1.27$\pm$0.23, 1.46$\pm$0.18 and 1.40$\pm$0.27, for SNL-0, SNL-1 and SNL-2, 
respectively. Note that the errors are larger by a factor of two than the assumed range of $\Upsilon_{\rm ref}$ in our default treatment,
implying that the 6dF data quality is not sufficient to {\it confirm} that the galaxies are `old', although they are certainly consistent with being so 
(the fitted ages are 10$\pm$3, 13$\pm$3 and 14$\pm$4 Gyr).

The possibility of younger ages leads to a tail in the probability distribution towards larger $\alpha$ (third panel of Figure~\ref{fig:alphapdf}).
The resulting mean $\alpha$ over the three galaxies is $\langle\alpha\rangle$\,=\,$1.19_{-0.13}^{+0.16}$  (Table~\ref{tab:alpha}, part d). 
(We account consistently for the different J-band solar absolute magnitude used in the CvD12a models, $M_J$\,=\,3.67, as compared to the M05 models used above.)
For all galaxies, the spectrum-fitting $\Upsilon_{\rm ref}$ is consistent with the previous assumptions, and hence yields $\alpha$ values similar to the 
default result, though with larger uncertainty. We note that the assumption of a single-burst star-formation history would lead to a slight overestimate of 
$\alpha$ using this method, if a galaxy actually harbours a more complex population: the V-band derived single-burst-equivalent age is {\it more} 
sensitive than the mass-to-light ratio to young stellar component, hence at fixed ``age'', a composite population has larger $\Upsilon_{\rm ref}$ than a single-burst.

Our results are subject to various sources of systematic errors in stellar population models.
Mancone \& Gonzalez (2012) report a total spread of $\sim$15\,per cent between different stellar
population models in the J-band mass-to-light ratio; of course the various models share several common ingredients and fundamental 
calibrations, so this does not necessarily reflect the full range of systematic uncertainty. 
One specific concern is that most stellar population models (including Maraston 2005 as used in our default calculation)
do not fully take into account the effects of non-solar element abundance patterns. 
Broadband colours can be affected directly by absorption features in the stellar atmosphere (especially in the blue) and by changes in the 
isochrones due to different opacities in the stellar interior.  Coelho et al. (2007) presented models taking both effects into account consistently, for 
a range of iron abundance Fe/H and $\alpha$ element enhancement. Comparing their 10\,Gyr SSP 
models we find that at fixed {\it total} metallicity (as assumed in M05), a 0.3\,dex increase in [$\alpha$/Fe] yields a $\sim$4\,per cent increase in 
J-band $\Upsilon_{\rm ref}$. Accounting for this effect would lead to a smaller derived $\alpha$. 

Finally, we note some more general sources of systematic error. 
Adopting $H_0$\,=\,67\,\kms\,Mpc$^{-1}$ from the {\it Planck} cosmology (Planck Collaboration 2014), rather than 
$H_0$\,=\,70\,\kms\,Mpc$^{-1}$ from WMAP7, would yield smaller $\alpha$ by $\sim$5\,per cent. 
Peculiar velocities of $\sim$500\,\kms\ would induce shifts of up to 5\,per cent in $\alpha$, 
depending on lens redshift. 
If the lens galaxies harbour very massive central black holes, then $\alpha$ would be a smaller than estimated by our method:
As a limiting case, if all lenses are assigned $M_{\rm BH}$\,=\,$10^{10}\,M_\odot$, comparable to the most-massive known 
objects (McConnell et al. 2011), $\langle\alpha\rangle$ is reduced to 0.94.

{\rev
Considering the range of values obtained above, we conclude that the dominant systematic uncertainty on $\langle\alpha\rangle$ arises from the choice
of whether to assume old ages or to fit $\Upsilon_{\rm ref}$ from 6dF spectra; the two methods yield results that differ by 13\,per cent. 
For clarity, we quote the results from both treatments in the 
abstract. The remaining systematics can be absorbed into an additional error component of $\pm$0.10, dominated by the variation among different
stellar population models, but also including the effect of different dark-matter assumptions, black hole contributions, abundance ratio effects, cosmology etc. 
No single factor seems likely to account for the $\sim$50\,per cent discrepancy between SNELLS and studies which indicate a heavy IMF in massive galaxies. 
}

\subsection{Field-of-view selection bias}\label{sec:fovbias}

\begin{figure}
\includegraphics[angle=270,width=85mm]{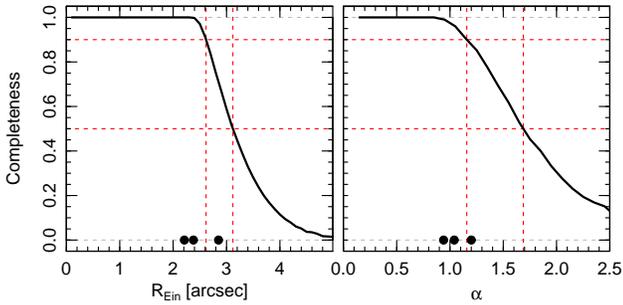}
\caption{Field-of-view completeness fraction, as a function of both Einstein radius, $R_{\rm Ein}$,  and IMF mass factor, $\alpha$. 
The latter assumes mass profiles informed by the 6dF luminosity profiles for $\sigma$\,$\ga$\,300\,\kms\ galaxies. 
The filled symbols at the bottom of the panels show $R_{\rm Ein}$ and $\alpha$ for SNL-0,1,2. Red dashed lines indicate the 90 and 50\,per cent limits.}
\label{fig:fovbias}
\end{figure}

The systematic errors reviewed in the previous section are uncertainties associated with measurements of $\alpha$ for the detected lenses.
Here we consider the more subtle effect of $\alpha$-dependent bias in selection of the lenses. This arises because the SINFONI field of view limits the observed
area around each target, so that large $R_{\rm Ein}$, and hence large $\alpha$, systems may be lost from the sample. 
In particular, for two-image systems like SNL-1 and SNL-2 (more common than quadruple images or rings which require closer 
alignment of lens and source), the more distant image may be formed at up to a maximum of 2\,$R_{\rm Ein}$. Although a second image will be formed 
at small radius and hence remain within the field of view, this is the demagnified 
image, so a large $R_{\rm Ein}$ system is less likely to be detected than a small $R_{\rm Ein}$ system.

To estimate the sample completeness as a function of $R_{\rm Ein}$, we used a singular isothermal sphere mass model, 
with a range in normalization, to compute image positions for a uniform grid of source locations, relative to the SNELLS 
observation geometry. All systems with $R_{\rm Ein}$\,$\la$\,2.4\,arcsec are recovered (i.e. both images within the field-of-view), 
after which the completeness drops with increasing $R_{\rm Ein}$: the recovered fraction $F(R_{\rm Ein})$ is 
90, 50 and 10\,per cent at $R_{\rm Ein}$\,=\,2.6, 3.1 and 4.1 arcsec respectively (Figure~\ref{fig:fovbias}, left).

The connection between $R_{\rm Ein}$ and  $\alpha$ depends on the mass profile of the galaxy, and hence estimating the completeness function
for $\alpha$ requires  a model for the galaxy-to-galaxy variation in the profiles. For this purpose, we use 
the half-light radius and surface brightness information for the 6dF survey galaxies  with $\sigma$\,$>$\,300\,\kms, from Campbell et al. (2014). 
We assume that the inner mass profile is dominated by the stellar mass, 
following an $R^{1/4}$ profile with a stellar mass-to-light ratio $\Upsilon$\,=\,$\Upsilon_{\rm ref}\alpha$\,=\,1.5$\alpha$, and allow a constant 
15\,per cent contribution of dark matter. For a given $\alpha$, the distribution of profiles leads to a distribution of $R_{\rm Ein}$, from which 
the completeness function can be derived.  As shown in the right-hand panel of Figure~\ref{fig:fovbias}, $F(\alpha)$ falls gradually from unity at 
$\alpha$\,=\,0.9, to 90, 50 and 10\,per cent at $\alpha$\,=\, 1.15, 1.70 and 2.70 respectively. 

Combining the completeness $F(\alpha)$ with an assumption for the underlying true distribution of $\alpha$, we can determine the extent of the bias 
in $\langle\alpha\rangle$. 
For a case tuned to resemble the dynamical results from Cappellari et al., with 
$\langle\alpha\rangle$\,=\,1.60  and scatter 0.3, the mean recovered $\alpha$ is 1.52, i.e. a bias of 0.08.
For lower input $\alpha$, where the completeness function is shallower, the bias is smaller. For the same assumed scatter of 0.3, our recovered
$\langle\alpha\rangle$\,=\,1.06 is compatible with a true underlying mean of 1.10 (Table~\ref{tab:alpha}, part e).

We conclude that although there is a selection bias against high-$\alpha$ lenses due to limited field-of-view, this effect is unlikely to account fully for
the small average value obtained from SNELLS, compared to results from other studies. 
{\rev Our final estimate for the underlying mean $\alpha$, correcting for the sample bias, is 
$\langle\alpha\rangle$\,=\,1.10$\pm$0.08$\pm$0.10 when the lenses are assumed to be old, and
$\langle\alpha\rangle$\,=\,$1.23^{+0.16}_{-0.13}\pm{0.10}$ when fitting ages from the 6dF spectra, where the second error indicates the
remaining systematic component.}

\section{Discussion}\label{sec:disc}

\begin{figure*}
\includegraphics[angle=0,width=175mm]{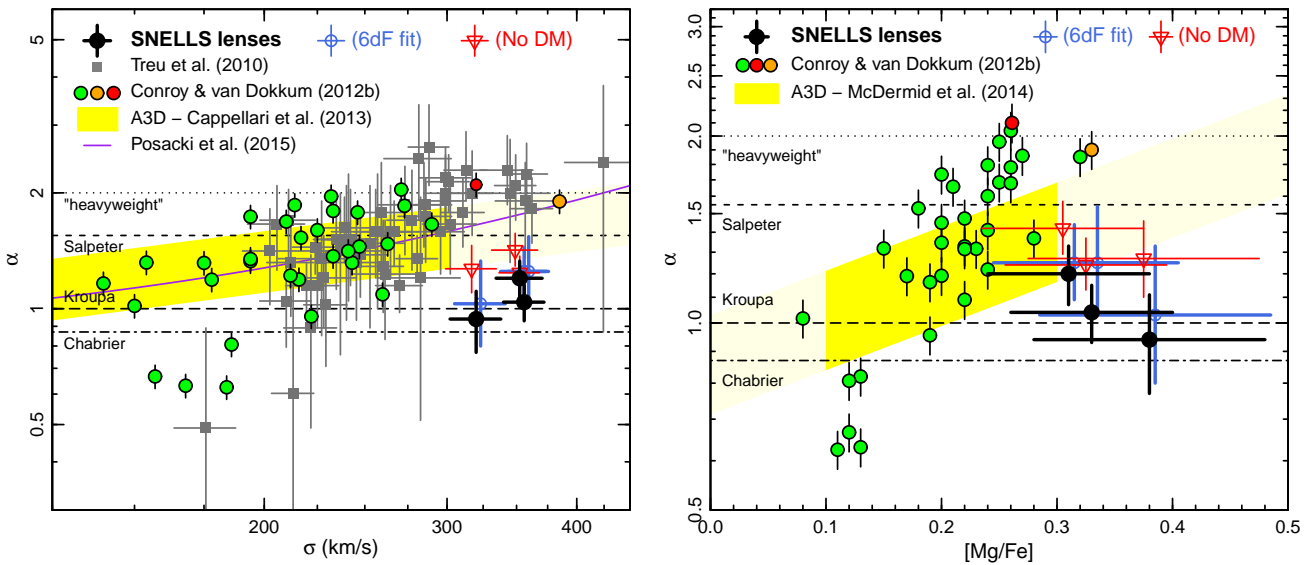}
\caption{The IMF $\alpha$ factor versus $\sigma$ and [Mg/Fe] from SNELLS, compared to previous results from spectroscopy, dynamics and distant lenses.
The SNELLS results using our default analysis are shown with filled black
symbols. Blue open points show the results when deriving $\Upsilon_{\rm ref}$ from the 6dF spectra; red triangles show the extreme assumption of zero dark matter.
 In the comparison data from Conroy \& van Dokkum (2012b), the orange
point is M87 and the red point is a stack of four Virgo cluster galaxies with high velocity dispersion, from Conroy \& van Dokkum (2010). 
For ATLAS3D, we show regions corresponding
to the fitted trends and estimated $\pm$1$\sigma$ intrinsic scatter  from Cappellari et al. (2013) and McDermid et al. (2014). The lighter yellow zones correspond to extrapolations
beyond the $\sigma$ and [Mg/Fe] ranges sampled by ATLAS3D. 
{\rev The purple line shows the quadratic fit to a combined ATLAS3D + SLACS sample by Posacki et al. (2015).}
Velocity dispersions have been  corrected to an aperture of $R_{\rm eff}/8$ throughout. 
}
\label{fig:alphacomp2}
\end{figure*}

Figure~\ref{fig:alphacomp2} presents the $\alpha$ estimates for the SNELLS lenses in the context of other recent studies, from spectroscopy 
(CvD12b), distant lenses (Treu et al. 2010, T10) and stellar dynamics (Cappellari et al. 2013, A3D). 
The figure is an updated version of that shown in SL13. 
We show $\alpha$ as a function of velocity dispersion and of the [Mg/Fe] abundance ratio (a proxy for star-formation timescale), the latter being
estimated from fits to the 6dF spectra {\rev using the CvD12a models}, as described in Section~\ref{sec:robalpha}.
Both quantities have been considered in previous work as possible predictors for IMF variations (CvD12b; Smith 2012; McDermid et al. 2014).

Figure~\ref{fig:alphacomp2} confirms that the  SNELLS galaxies have lensing-derived $\alpha$ factors that are smaller than the average values derived by the other
studies for galaxies of similar properties\footnote{Note that the $\alpha$-vs-$\sigma$ relation could in principle be affected by a subtle lensing selection bias: 
lensing is most effective for galaxies with long axis aligned towards the observer, maximizing projected mass. 
While $\alpha$ is not biased by this effect (since the projected luminosity
is also increased), the velocity dispersion for such galaxies would be biased high, relative to a random orientation. The [Mg/Fe] abundance ratio is of course not affected 
by such a bias. The high [Mg/Fe] for our galaxies, which lie on the fairly tight [Mg/Fe]--$\sigma$ relation, suggests they do in fact have intrinsically large velocity dispersions.}.
In this section, we discuss the comparison with each of the other studies shown in the figure, and also comment on compactness as a possible predictor for IMF variation.

\subsection{Comparison with ATLAS3D}

{\rev

For the comparison with ATLAS3D (A3D) dynamical estimates of $\alpha$, we show the correlation with $\sigma$  as reported by 
Cappellari et al. (2013) and the trend with [Mg/Fe] from McDermid et al. (2014), with the estimated intrinsic scatter estimated in those papers.

The A3D sample does not include any galaxies with $\sigma$\,$>$\,300\,\kms, but an extrapolation of their trend with $\sigma$ 
would indicate $\alpha$\,=\,1.60 at the mean $\sigma$ of the SNELLS galaxies (345\,\kms\ at $R_{\rm eff}/8$, 
equivalent to $\sigma_{\rm e}$\,=\,315\,\kms). Taking into account their estimated intrinsic scatter of 0.084\,dex\footnote{\rev For these
comparisons, we assume that the scatter around the A3D trend is gaussian in $\log\alpha$.},
the observed mean of 1.10$\pm$0.08 (for the default, ``assumed old" results, corrected for field-of-view bias, but neglecting systematic errors for now) deviates
from the A3D relation by $\Delta$\,=\,2.8$\sigma$, and Pr($\Delta$)=\,0.998.  
Including the systematic error component of $\pm$10\,per cent in $\alpha$
reduces the offset of the mean to 2.3$\sigma$; this is a lower limit, because some of the systematic error sources are common to both methods. 
A $\chi^2$ test, which does not take account of the direction of offsets from the trend, does not detect such a significant disagreement. 
Table~\ref{tab:a3dcomp} summarizes this comparison, together with equivalent tests for other $\alpha$--$\sigma$ relations quoted
by Cappellari et al., as well as by Posacki et al. (2015) using a combined A3D+SLACS sample.
We conclude that the {\it size} of discrepancies from the A3D $\alpha$--$\sigma$ trend are not especially surprising; however, when
taking into account that all three galaxies are offset in the same direction, there is tension at the 2--3\,$\sigma$ level.

At the average [Mg/Fe] for the SNELLS lenses (0.34), the McDermid et al. trend predicts $\alpha$\,=\,1.50 with intrinsic
dispersion of 0.076\,dex. Hence the SNELLS mean $\alpha$ is discrepant from the A3D $\alpha$--[Mg/Fe] relation at a similiar $\sim$2.5$\sigma$ level.

We conclude that although our results significantly favour a Milky-Way-like $\alpha$ over a Salpeter-like value for {\it all} galaxies, the present SNELLS sample is only marginally inconsistent with the A3D results, when allowing for the substantial intrinsic scatter around their trends. 
}

\begin{table*} 
\caption{\rev Comparison of SNELLS results against the trends reported by Cappellari et al. (2013) for A3D and 
Posacki et al. (2015) for a combined A3D  and SLACS sample.
The trends are taken from figure 13 (bottom and top panels) of Cappellari et al. and equations 5 and 6 of Posacki et al. (2015).
For each of the proposed $\alpha$--$\sigma$ relations, $\langle\log(\alpha)\rangle$ is the average $\log\alpha$ predicted
at the mean velocity dispersion of the SNELLS sample and $\delta$ is the {\it intrinsic} scatter around the trend. 
The  following two columns summarise a $\chi^2$ test for the three SNELLS observations with respect to the given trend,
accounting for measurement error in $\alpha$ and $\sigma$, field-of-view (FoV) bias, and the intrinsic dispersion $\delta$.
(We use $\delta$\,=\,0.081 for intrinsic scatter in the the Posacki et al. trends, based on private communication from the authors.)
The final two columns are for a comparison of the mean: $\Delta$ is the deviation of measured mean $\log\alpha$
from the prediction, in units of the error again including field-of-view bias and intrinsic dispersion $\delta$.
 } \label{tab:a3dcomp}
\begin{tabular}{lcccccccccccccc} 
\hline
Trend &   $\langle\log(\alpha)\rangle$ & $\delta$ & \ \  &  $\chi^2$ & Pr($\chi^2$) & $\Delta$ & Pr$(\Delta$) \\
\hline
SNELLS (assumed old; FoV-corrected) & 0.041$\pm$0.031 \\
Cappellari et al. (2013) ``preferred" & 0.203$\pm$0.048 & 0.084 & &  8.26 & 0.9591 & $-2.84$ & 0.9977 \\
Cappellari et al. (2013) ``all galaxies'' & 0.167$\pm$0.048 & 0.083 & & 5.18 & 0.8406 & $-2.22$ & 0.9869 \\
Posacki et al. (2015) linear  & 0.206$\pm$0.047 & 0.081 & & 9.04 & 0.9711 & $-2.98$ & 0.9986 \\
Posacki et al. (2015) quadratic  & 0.234$\pm$0.047 & 0.081 & & 11.21 & 0.9933 & $-3.47$ & 0.9997 \\
\hline
\end{tabular}
\end{table*}

\subsection{Comparison with SLACS}

At face value, the most surprising discrepancy in Figure~\ref{fig:alphacomp2} is that between SNELLS and the results of T10 from SLACS, which is also based on strong lensing.
In particular, for twelve galaxies with $\sigma$\,$>$\,300\,\kms\ (after approximate correction to the the same aperture $R_{\rm eff}$/8 as used by A3D and CvD12), T10
find $\langle\alpha\rangle$\,$\approx$\,2.0 with no measurable intrinsic scatter. 

Our methods most clearly depart from the approach taken by T10 in the treatment of dark matter contributions. 
T10 fit a spherical two-component (dark plus stellar) mass model to the measured velocity dispersion and Einstein radius.
The velocity dispersion
depends mainly on the {\it enclosed} mass at small radius, while $R_{\rm Ein}$ constrains the {\it projected} mass at larger radius. These observables 
are employed together to disentangle the contributions of the two components of the mass model.
In the T10 analysis, the dark matter is assumed to follow the Navarro et al. (1995) density profile, 
i.e. with $\rho(r)$\,$\sim$\,$r^{-1}$ in the region sampled by the data. 
{\rev Posacki et al. (2015) show that essentially identical results are obtained using more general axisymmetric  
models constrained by the observed light distribution, velocity dispersion and Einstein-aperture projected mass.}

In SNELLS we have instead chosen to assume a dark matter contribution informed by  results from the EAGLE 
hydrodynamical simulation. To test whether this difference in method accounts for the divergent results obtained, we have used the 
$R_{\rm Ein}$ tabulated by Auger et al. (2009) to derived the projected dark-matter mass predicted 
for the SLACS lenses by our method. 
For the twelve galaxies with the highest velocity dispersions, we find that the dark-matter masses predicted from EAGLE are larger, 
by a median factor of 2.6 than those recovered by T10 from their two-component mass models. 
Using the EAGLE approach instead would increase the median dark-matter fraction (within $R_{\rm Ein}$) 
in the SLACS galaxies from 14\,per cent to 33\,per cent, and reduce the median IMF mass factor from 2.0 to 1.6, for these high-$\sigma$
galaxies. (Across the whole SLACS sample, 
the median $\alpha$ is reduced from 1.6 to 1.3 with the imposition of EAGLE dark-matter content.) 
This reduction in $\alpha$ is similar to what was obtained by Auger et al. (2010) from a re-analysis of the SLACS data, allowing explicitly for halo contraction effects.
In particular, when modelling contraction following the Gnedin et al. (2004) prescription, Auger et al. recover a shallower trend with galaxy mass, 
reaching a maximum of $\alpha$\,$\approx$\,1.4.

A second methodological difference between the studies is in the method for assigning the reference mass-to-light ratio, $\Upsilon_{\rm ref}$: T10 use
estimates from fits to the spectral energy distribution (SED) based on broad-band optical and (sometimes) IR data, described by Auger et al. (2009). In fitting the SEDs,
however, they impose an informative prior on the star-formation histories restricting the onset of star-formation to early epochs (1\,$<$\,$z$\,$<$\,5, uniform in time) and 
favouring short star-formation timescales ($\tau$\,$\la$1\,Gyr). In practice these priors force the lens galaxies to have old stellar populations, and hence the SLACS 
$\Upsilon_{\rm ref}$ is within $\sim$10\,per cent of what would be imposed by our default assumption for SNELLS. 

{\rev We have not found a satisfying explanation for the factor-of-two disagreement between the lensing-based estimates of $\alpha$ from 
SNELLS and from SLACS for similarly-high-$\sigma$ ellipticals. Halo contraction could account for part of the 
difference\footnote{\rev Since the SLACS analysis incorporates the velocity dispersion, which responds to the {\it enclosed} mass, it is more sensitive to 
halo contraction than the SNELLS method which uses only projected quantities.}, but analyses of SLACS and other datases appear to favour uncontracted
NFW halos (e.g. Dutton et al. 2013; Dutton \& Treu 2014, but see Grillo 2012; Sonnenfeld et al. 2012). 
While it is possible that our dark-matter corrections from EAGLE could have been overestimated, we note again that even assuming
no dark matter at all, the mean $\alpha$ from SNELLS is still inconsistent with the average of $\sim$2 derived for SLACS at $\sigma$\,$>$\,300\,\kms.
}

\subsection{Comparison with CvD12b}

As emphasised by Smith (2014), there are important differences between the A3D and CvD12b results for $\alpha$, both at the level of individual galaxies, and in terms 
of trends with respect to $\sigma$ and [Mg/Fe]. In particular, in CvD12b, $\alpha$ is more strongly correlated than with $\sigma$, while the reverse  holds for A3D. 
For galaxy properties similar to those of the SNELLS lenses, CvD12b find larger values of $\alpha$ than A3D, 
with an average of $\sim$2, and hence the apparent discrepancy between CvD12b and SNELLS is $\sim$4\,$\sigma$ (assuming same intrinsic scatter as for A3D).
Since publication of CvD12b, Conroy \& van Dokkum (in preparation) have expanded their observational sample, and made numerous improvements to the spectral 
fitting method. These updates have led to somewhat smaller derived $\alpha$ factors, on average, for the most massive galaxies. 
More detailed comparisons should await publication of the revised spectroscopic constraints.

\subsection{Compactness as a predictor for IMF variations?}

Aside from velocity dispersion and [Mg/Fe], another property of galaxies which has been proposed as being related to IMF variation is  
``compactness''. Since compact massive galaxies are abundant at high redshifts but rare in the local universe (e.g. Trujillo et al. 2006), 
it has been argued that low-redshift compact galaxies are surviving ``pristine'' relics of the early violent starburst epoch,
where non-standard IMFs may have been imposed (L\"asker et al. 2013; Chabrier, Hennebelle \& Charlot 2014).  
In SL13, we argued that a dependence on compactness might account for the 
difference between the ``lightweight'' IMF in SNL-0 and the heavy IMF found by L\"asker et al. in the $z$\,=\,0.1 compact galaxy ``b19''.

\begin{figure}
\includegraphics[angle=270,width=85mm]{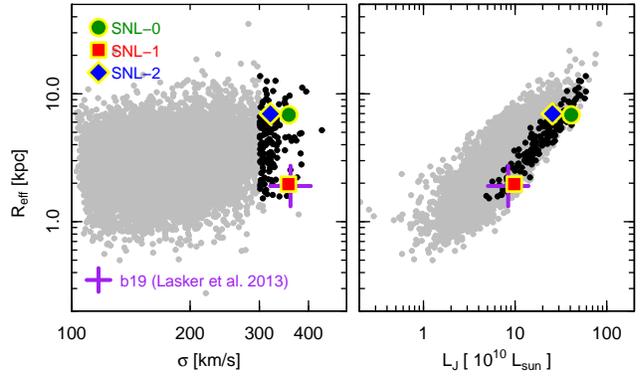}
\caption{
{\rev Velocity dispersion, $\sigma$, effective radius $R_{\rm eff}$ and total luminosity $L_J$ for  the three SNELLS lenses,
compared to the full 6dF galaxy sample. Dark points show all 6dF galaxies with $\sigma$\,$>$\,300\,\kms. The three targets span 
most of the range in size present  among high-$\sigma$ galaxies. For comparison, we show also the compact galaxy discussed by 
L\"asker et al. (2013), which has structural properties similar to those of SNL-1.
}}
\label{fig:sixdffp}
\end{figure}

{\rev The addition of the two new SNELLS lenses does not support this speculation. 
Figure~\ref{fig:sixdffp} shows projections of the 6dF Fundamental Plane from Campbell et al. (2014), and the location of the three 
SNELLS lenses for comparison. Although the sample size remains small, the lenses  acceptably reproduce the distribution of size and luminosity for
defined by the whole $\sigma$\,$>$\,300\,\kms\ galaxy population. SNL-0 and SNL-2 are somewhat larger and more luminous than average, while
SNL-1 lies among the most compact of the high-$\sigma$ galaxies. The properties of SNL-1, where lensing requires
$\alpha$\,$\approx$\,1.2 are in fact very similar to those of b19,
where L\"asker et al. find $\alpha$\,=\,2.2$\pm$0.6 from dynamics (unless b19 harbours an extremely massive central black hole).

In summary, we do not find support from SNELLS for a strong correlation of $\alpha$ with compactness among the $\sigma$\,$>$\,300\,\kms population.
Clearly this conclusion is heavily dependent on the results for SNL-1, however, and the current small sample cannot rule out a correlation having large 
intrinsic scatter. }

\section{Summary}\label{sec:concs}

We have presented first results from the SNELLS survey for nearby strong-lensing ellipticals, demonstrating the success of our
infra-red IFU search method with the discovery of two new $z$\,$\la$\,0.05 lenses, as well as recovering the only previously-known
case, ESO325--G004. The survey yield, i.e. identified lenses per new observed candidate, is broadly compatible with the $\sim$10\,per cent rate
estimated from the H$\alpha$ luminosity function of Sobral et al. (2013) (but note that one of the new sources, like that in ESO325--G004, is  
an [O\,{\sc iii}] emitter, and hence not included in our calculations).
Compared to distant lens samples, SNELLS selects systems in which the Einstein radius is a smaller
fraction of the stellar effective radius, so that the lensing mass is more reliably dominated by the stellar component. 

Our lens galaxies have high velocity dispersions ($\sigma$\,$>$\,300\,\kms) and $\alpha$-enhanced abundances ([Mg/Fe]\,$\ga$\,0.3), 
and hence are characteristic of the systems where previous works have inferred heavy IMFs though dynamical (Cappellari et al. 2013), lensing (Treu et al. 2010)
or spectroscopic (Conroy \& van Dokkum 2012b) analyses. 
Notwithstanding these properties, the SNELLS galaxies have lensing-derived stellar mass-to-light ratios that are individually or collectively 
compatible with a Milky-Way-like Kroupa (2001) IMF. A heavier-than-Salpeter IMF (without intrinsic scatter) is disfavoured {\rev (3.5$\sigma$ significance)}
if we assume the stellar populations to be old {\rev ($\langle\alpha\rangle$\,=\,1.10$\pm$0.08$\pm$0.10)}, but cannot be excluded if we fit the ages from the 
currently-available spectra {\rev ($\langle\alpha\rangle$\,=\,1.23$\pm$0.15\,$\pm$0.10)}. {\rev Very heavy IMFs with $\alpha$\,$>$\,2 are firmly rejected using either approach.}
The result for ESO325--G004 (SNL-0) is consistent with our previous estimates (Smith \& Lucey 2013), despite several changes in data and methods.
The mean IMF result obtained from SNELLS is mildly inconsistent (at the 2--3$\sigma$ level) with the IMF-versus-$\sigma$ trend reported by 
Cappellari et al. (2013), after accounting for intrinsic scatter and our field-of-view selection bias. There is a larger discrepancy with respect to the 
Treu et al. (2010) analysis of the SLACS lens sample, part of which can be attributed to the different treatment of projected dark-matter contributions.
The SNELLS results are also apparently in conflict with the bottom-heavy IMFs inferred from spectroscopic fitting by Conroy \& van Dokkum (2012b) at similarly 
high $\sigma$ and Mg/Fe. 

The largest source of uncertainty in each {\it individual} measurement of $\alpha$ is in the treatment of $\Upsilon_{\rm ref}$, while the extent to which we
can infer the {\it distribution} of $\alpha$ in the population of giant early-type galaxies is limited by the small sample size. 
In future work with SNELLS, we plan to observe more lens candidates, with the goal of doubling the sample of low-redshift 
lensing systems available for analysis. We are also pursuing homogeneous high-signal-to-noise optical/IR spectroscopy of the lens galaxies. The optical spectra 
will yield more reliable estimates for the stellar age and metallicity of the lenses, and hence $\Upsilon_{\rm ref}$,  obviating the need to assume purely old 
stellar populations. The red and IR spectra will cover a range of IMF sensitive features (including the Na\,{\sc i} 8200-\AA\ doublet and Fe\,H Wing--Ford band) 
providing a stringent test of spectroscopic models in cases where the low-mass stellar content is constrained by the lensing M/L. 
Eventually, we will to use these systems to compare multiple methods of IMF estimation (lensing, dynamical, spectroscopic) for a common sample of 
nearby galaxies, where high quality data can be readily obtained.

At present, SINFONI is the only IFU operating with sufficiently wide contiguous field-of-view to perform a study of this kind efficiently in the IR. 
However, the success of our search strategy suggests it should be worthwhile to search for lensed background sources in large optical 
IFU datasets, e.g. from the SAMI (Allen et al. 2015) and MaNGA (Bundy et al. 2015) surveys.
MaNGA is the most promising of these options, since its wider spectral coverage (to 1\,$\mu$m) will detect 
background H$\alpha$ sources out to $z$\,$\approx$\,0.5 and [O\,{\sc ii}] to $z$\,$\approx$\,1.5. 
In the longer term, the HARMONI first-light instrument (Thatte et al. 2014) 
on the 39m European Extremely Large Telescope, with 5$\times$10\,arcsec$^2$ field-of-view IFU,
should enable such deep observations that strongly-lensed background line emitters could be detected for {\it any} chosen massive elliptical.

\section*{Acknowledgements}
RJS was supported by the STFC Durham Astronomy Consolidated Grant 2014--2017 (ST/L00075X/1). 
We are grateful to Matthieu Schaller, Richard Bower and the EAGLE consortium for help in deriving estimates 
of the projected dark matter masses in massive galaxies, and to Ian Smail for useful discussions during development of the project.
We thank an anonymous referee for helpful comments and suggestions.
This research has made use of the NASA/IPAC Extragalactic Database (NED) which is operated by the Jet Propulsion Laboratory, 
California Institute of Technology, under contract with the National Aeronautics and Space Administration.
This publication makes use of data products from the Two Micron All Sky Survey, which is a joint project of the University of Massachusetts and the 
Infrared Processing and Analysis Center / California Institute of Technology, funded by the National Aeronautics and Space Administration and the National Science Foundation. The data used in this work are available through the ESO science archive under programme IDs 093.B-0193 and 293.B-5026, 
and through databases for 2MASS (www.ipac.caltech.edu/2mass) and the 6dF Galaxy Survey (www.6dfgs.net)

{}

\label{lastpage}

\end{document}